\documentclass{article}

\usepackage{amsfonts}
\usepackage{amsmath}
\usepackage{indentfirst,latexsym,bm}
\usepackage{amssymb}
\usepackage{amsmath}
\usepackage{graphics}
\usepackage{graphicx}
\usepackage{amsfonts}
\usepackage{amsmath}
\usepackage{indentfirst,latexsym}
\usepackage{amssymb}
\usepackage{amsmath}
\usepackage{graphics}
\usepackage{authblk}
\usepackage{color}
\usepackage{float}
\usepackage{latexsym}
\usepackage{booktabs}
\usepackage{multirow}
\usepackage{graphicx}
\usepackage{epsfig}
\usepackage{geometry}
\usepackage{amscd}
\usepackage{lscape}
\usepackage{float}
\usepackage{amsmath}
\usepackage{amssymb}
\usepackage{tabularx}
\usepackage{enumerate}
\usepackage{amsfonts}
\usepackage{mathrsfs}
\usepackage{eufrak}
\usepackage{graphicx}
\usepackage{dsfont}
\usepackage{epsfig}
\usepackage[ruled, lined, vlined, shortend, algosection]{algorithm2e}
\usepackage{longtable}
\usepackage{pdflscape}
\usepackage{lscape}
\usepackage{curves}
\usepackage{epic}
\usepackage{amsfonts}
\usepackage{amsmath}
\usepackage{indentfirst,latexsym,bm}
\usepackage{amssymb}
\usepackage{amsmath}
\usepackage{graphics}
\usepackage{authblk}
\usepackage{color}
\usepackage{float}
\usepackage{latexsym}
\usepackage{booktabs}
\usepackage{multirow}
\usepackage{graphicx}
\usepackage{epsfig}
\usepackage{geometry}
\usepackage{amscd}
\usepackage{lscape}
\usepackage[dvipdfm]{hyperref}

\numberwithin{equation}{section}

\newtheorem{theorem}{Theorem}[section]

\newtheorem{assumption}[theorem]{Assumption}

\newtheorem{corollary}[theorem]{Corollary}

\newtheorem{definition}[theorem]{Definition}

\newtheorem{lemma}[theorem]{Lemma}

\newtheorem{proposition}[theorem]{Proposition}

\newenvironment{proof}[1][Proof]{\textbf{#1.} }{\ \rule{0.5em}{0.5em}}
\DeclareMathOperator*{\esssup}{ess\,sup}

\DeclareMathOperator*{\essinf}{ess\,inf}

\begin{document}

\title{\textbf{A Multidimensional Exponential Utility Indifference Pricing Model
with Applications to Counterparty Risk}\footnotetext{We
thank the Oxford-Man Institute for partial support of this research. 
We thank participants at the 3rd AQFC conference (Hong Kong, July, 2015) 
and also Damiano Brigo, Rama
Cont, David Hobson, Monique Jeanblanc, Lishang Jiang, Eva
L\"utkebohmert, Andrea Macrina, Shige Peng, Martin Schweizer, Xingye
Yue, and Thaleia Zariphopoulou for their helpful discussions.
}}

\author[$\dag$,$\sharp$]{\small{\textsc{Vicky Henderson}}}

\author[$\ddag$,$\sharp$]{\small{\textsc{Gechun Liang}}}

\affil[$\dag$]{\small{\textsc{Department of Statistics, University
of Warwick, Coventry, CV4 7AL, U.K.}}\\

\texttt{Vicky.Henderson@warwick.ac.uk}}

\affil[$\ddag$]{\small{\textsc{Department of Mathematics, King's College London, London, WC2R 2LS, U.K.}}\\

\texttt{gechun.liang@kcl.ac.uk}}

\affil[$\sharp$]{\small{\textsc{Oxford-Man Institute, University of
Oxford, Oxford, OX2 6ED, U.K.}}}

\maketitle

\begin{abstract}
This paper considers exponential utility indifference pricing for a
multidimensional non-traded assets model subject to inter-temporal
default risk, and provides a semigroup approximation for the utility
indifference price. The key tool is the splitting method, whose
convergence is proved based on the Barles-Souganidis monotone scheme,
and the convergence rate is derived based on Krylov's shaking the
coefficients technique. We apply our methodology to study the
counterparty risk of derivatives in incomplete markets.
\\

\noindent\textit{Keywords}: Utility indifference pricing, \and
reaction-diffusion PDE with quadratic gradients, \and splitting
method, \and monotone scheme, \and shaking the coefficients
technique,
\and counterparty risk.\\

\noindent\textit{Mathematics Subject Classification (2010)}: 91G40,
\and 91G80, \and 60H30.
\end{abstract}


%
\newpage
\section{Introduction}

The purpose of this article is to consider \emph{exponential}
utility indifference pricing in a multidimensional non-traded assets
setting subject to intertemporal default risk, which is motivated
by our study of counterparty risk of derivatives in incomplete
markets. Our interest is in pricing and hedging derivatives written
on assets which are not traded. The market is incomplete as the
risks arising from having exposure to non-traded assets cannot be
fully hedged. We take a utility indifference approach whereby the
utility indifference price for the derivative is the cash amount the
investor is willing to pay such that she is no worse off in expected
utility terms than she would have been without the derivative.

There has been considerable research in the area of {\it
exponential} utility indifference valuation, but despite the
interest in this pricing and hedging approach, there have been
relatively few explicit formulas derived. The well known {\it one
dimensional non-traded assets model} is an exception and in a
Markovian framework with a derivative written on a single non-traded
asset, and partial hedging in a financial asset, Henderson and
Hobson \cite{Henderson2}, Henderson \cite{Henderson1}, and Musiela
and Zariphopoulou \cite{Musiela} used the Cole-Hopf transformation
(\emph{or distortion power}) to linearize the non-linear PDE for the
value function. This trick results in an explicit formula for the
exponential utility indifference price. Subsequent generalizations
of the model from Tehranchi \cite{Teh}, Frei and Schweizer
\cite{FSI} and \cite{Schweizer} showed that the exponential utility
indifference value can still be written in a closed-form expression
similar to that known for the Brownian setting, although the
structure of the formula can be much less explicit. On the other
hand, Davis \cite{Davis} used the duality to derive an explicit
formula for the optimal hedging strategy (see also Monoyios
\cite{Monoyis}), and Becherer \cite{Bec0} showed that the dual
pricing formula exists even in a general semimartingale setting.

As soon as one of the assumptions made in the one dimensional
non-traded assets model breaks down, explicit formulas are no longer
available. For example, if the option payoff depends also on the
traded asset, Sircar and Zariphopoulou \cite{MR2124276} developed
bounds and asymptotic expansions for the exponential utility
indifference price. In an energy context, we may be interested in
partially observed models and need filtering techniques to
numerically compute expectations (see Carmona and Ludkovski
\cite{CL} and Chapter 7 of \cite{MR2547456}). If the utility
function is not exponential, Henderson \cite{Henderson1} and Kramkov
and Sirbu \cite{KS} developed expansions in small quantity for the
utility indifference price under power utility.


In this paper, we study exponential utility indifference valuation
in a multidimensional setting subject to intertemporal default risk
with the aim of developing a pricing methodology.
The main economic motivation for us to develop the multidimensional
framework is to consider the counterparty default risk of options
traded in over-the-counter (OTC) markets, often called
\emph{vulnerable options}. The credit crisis has brought to
the forefront the importance of counterparty default risk as numerous high profile defaults lead to counterparty losses.
In response, there have been many recent studies (see, for example,
Bielecki et al \cite{BCJZ} and Brigo et al \cite{Brigo-MF})
addressing in particular the counterparty risk of credit default
swaps (CDS). In contrast, there is relatively little recent work on
counterparty risk for other derivatives, despite OTC options being a
sizable fraction of the OTC derivatives market.\footnote{In fact,
OTC options comprised about 10\% of the \$600 trillion (in terms of
notional amounts) OTC derivatives market at the end of June 2010
whilst the CDS market was about half as large at around \$30
trillion.} The option holder faces both price risk arising from the
fluctuation of the assets underlying her option and counterparty
default risk that the option writer does not honor her obligations.
Default occurs either when the assets of the counterparty are below
its liabilities at maturity (\emph{the structural approach}) or when
an exogenous random event occurs (\emph{the reduced-form approach}),
so intertemporal default is considered. In our setting, the assets
of the counterparty and the assets underlying the option may be
non-traded and thus a multidimensional non-traded assets model
naturally arises.

Our use of the utility indifference approach is motivated by its
recent use in credit risk modeling where the concern is the default
of the reference name rather than the default of the counterparty.
Utility based pricing has also been utilized by Bielecki and
Jeanblanc \cite{Bielecki1}, Sircar and Zariphopolou \cite{MR2642963}
and recently Jiao et al \cite{Jiao1} \cite{Jiao2} in an intensity
based setting. Several authors have applied it in modeling of
defaultable bonds where the problem remains one dimensional, see in
particular Leung et al \cite{leung}, Jaimungal and Sigloch
\cite{JS2009}, and Liang and Jiang \cite{LJ2009}. In contrast,
options subject to counterparty risk are a natural situation where
two or more dimensions arise.


Our first contribution is the derivation of a reaction-diffusion
partial differential equation (PDE) in Theorem
\ref{Markov_lemma_111} to characterize the utility indifference
price in our multidimensional setting, where we do not rely on the
dynamic programming principle. Instead, we consider the
associated utility maximization problems for utility indifference
valuation from a risk-sensitive control perspective. We first
transform our utility maximization problems into risk-sensitive
control problems, and employ the comparison principle to derive a
quadratic backward stochastic differential equation (BSDE)
representation for the utility indifference price. See also Theorem
2.3 of Henderson and Liang \cite{HL} where we derived a quadratic
BSDE representation for the utility indifference price in a
non-Markovian setting with full recovery rate. The pricing PDE is
then obtained by an application of nonlinear Feynman-Kac formula for
quadratic BSDE (see Section 3 of Kobylanski \cite{Kobylanski}).

Our main contribution is to develop a semigroup approximation for
the pricing PDE by the splitting method.
In our multidimensional setting, the Cole-Hopf transformation (as in
the one dimensional model) cannot be applied directly since the
coefficients of the quadratic gradient terms do not match, and due
to the existence of the intertemporal payment term. Motivated by
the idea of the splitting method (\emph{or fractional step,
prediction and correction, etc}) in numerical analysis, we split the
pricing equation into two semilinear PDEs with quadratic gradients
and an ordinary differential equation (ODE) with Lipschitz
coefficients, such that the Cole-Hopf transformation can be applied
to linearize both PDEs, and the Picard iteration can be used to
linearly approximate the ODE.

The idea of splitting in our setting is as follows. The time
derivative of the pricing equation depends on the sum of semigroup
operators corresponding to different factors. For each subproblem
corresponding to each semigroup there might be an effective way
providing solutions, but for the sum of the semigroups, there may
not be an accurate method. The application of the splitting method
means that we treat the semigroup operators separately. We prove
that when the mesh of the time partition goes to zero, the
approximate price will converge to the utility indifference price in
Theorem \ref{semigrouptheorem}, relying on the \emph{monotone
scheme} introduced by Barles and Souganidis \cite{Barles}: any
\emph{monotone, stable} and \emph{consistent} numerical scheme
converges to the correct solution provided there exists a comparison
principle for the limiting equation. Moreover, by employing the
\emph{shaking the coefficients technique} introduced by Krylov
\cite{Krylov} (see also Barles and Jacobsen \cite{BJ,Jakobsen} for
its development), we are able to obtain the convergence rate of our
splitting algorithm in Theorems \ref{theorem_error_1} and
\ref{theorem_error_2}. The key to apply the monotone scheme and
shaking the coefficients technique is to derive an consistent error
estimate, which we prove by utilizing a coupled forward backward
stochastic differential equation (FBSDE) representation for the
solution of PDEs in (\ref{semigroupequation}).


Our third contribution is the application of the splitting method to
compute prices of derivatives on a non-traded asset and where the
derivative holder is subject to non-traded counterparty default
risk. In contrast to the complete market Black-Scholes style
formulas obtained by Johnson and Stulz \cite{Johnson}, Klein
\cite{Klein} and Klein and Inglis \cite{KI2001}, we show the
significant impact that non-tradeable risks have on the valuation of
vulnerable options and the role played by partial hedging. In
particular, our numerical illustrations quantify the effect of
non-tradeable price and default risk on option prices. The magnitude
of the potential discounts to the complete market price depends upon
the likelihood of default and if default is likely, the discount can
be extremely high. We also observe that put values can decrease with
maturity (in absence of dividends) in situations where the risk of
default is significant.

Splitting methods have been used to construct numerical schemes for
PDEs arising in mathematical finance (see the review of Barles
\cite{Barles-review}, and Tourin \cite{Tourin} with the references
therein). Recently, Nadtochiy and Zariphopoulou \cite{NZ} applied
splitting to the {\it marginal} Hamilton-Jacobi-Bellman (HJB)
equation arising from optimal investment in a two-factor stochastic
volatility model with general utility functions. They show their
scheme converges to the unique viscosity solution of the limiting
equation. Whilst they also apply splitting to an incomplete market
problem, their focus is to deal with the lack of a verification
theorem in their setting. In contrast, we propose a multidimensional
model subject to intertemporal default risk with time and space
variable coefficients and exponential utility. Our contributions
include proposing a splitting approach for utility indifference
pricing in a multidimensional non-traded assets model with
intertemporal default risk, identifying how to split the resulting
pricing PDE, and moreover, we prove the convergence rate of our
splitting method by using advanced techniques from the theory of
viscosity solutions. Other recent works of Halperin and Itkin
\cite{HI1, HI2} propose the use of splitting methods to price
options on a single illiquid bond via mixed static-dynamic hedging.
Our model is instead designed for multiple non-traded assets, which
is necessary for our treatment of
counterparty risk in a hybrid structural-reduced form setting. Finally, Tan \cite{Tan}
proposes a splitting method for fully nonlinear degenerate parabolic PDEs and applies it to Asian options and commodity trading. 

The paper is organized as follows: In Section 2, we present our
multidimensional exponential utility indifference pricing model, and
propose a splitting method to solve the pricing equation. The
convergence of the splitting algorithm and its convergence rate are
proved in Section 2.3 and 2.4, respectively. In Section 3, we apply
the method to study counterparty risk. We conclude with the
verification theorem of the pricing equation in the Appendix.


\section{A Splitting Method for Utility Indifference Valuation} \label{sec-model}


\subsection{Model Setup}
Let ${\cal{W}}=(W^1,\dots,W^{n+2})$ be an $(n+2)$-dimensional
Brownian motion on a filtered probability space
$(\Omega,\mathcal{F}, \mathbb{F}=\{{\mathcal{F}}_t\}_{t\geq 0}
,\mathbf{P})$ satisfying the \emph{usual conditions}, where
$\mathcal{F}_t$ is the augmented $\sigma$-algebra generated by $
({\cal{W}}_u:0 \leq u \leq t)$. The market consists of a risk-free
bank account with price $1$, a set of observable but non-traded
assets $\mathcal{S}=(S^1,\dots,S^n)$, whose \emph{logarithm} price
processes are driven by
\begin{equation}\label{dynamics of nontraded assets_2}
{dS_t^{i}}=\mu_i(\mathcal{S}_t,t)dt+\sigma_{i}(\mathcal{S}_t,t)dW_t^i+\bar{\sigma}_{i}(\mathcal{S}_t,t)dW_t^{n+1}
\end{equation}
for $i=1,\dots,n$, and a traded financial index $P$, whose price
process is driven by
\begin{equation}\label{P_equ_2}
\frac{dP_t}{P_t}=\mu_Pdt+\bar{\sigma}_{P}dW_t^{n+1}+\sigma_{P}dW_t^{n+2}.
\end{equation}
The price of each non-traded asset $S^i$ reflects exposure to the
traded or market risk $W^{n+1}$ through volatility
$\bar{\sigma}_i(\mathcal{S}_t,t)$ and non-traded idiosyncratic risk
$W^{i}$ through idiosyncratic or undiversifiable volatility
$\sigma_i(\mathcal{S}_t,t)$. We define the following parameters for
the financial index $P$:
\begin{align*}
\theta^P=\frac{(\mu_P)^2}{(\sigma_{P})^2};\ \ \ &\bar{\theta}^P=\frac{(\mu_P)^2}{(\sigma_{P})^2+(\bar{\sigma}_{P})^2};\\
\vartheta^P=\frac{\mu_P\bar{\sigma}_{P}}{(\sigma_{P})^2};\ \ \ &\bar{\vartheta}^P=\frac{\mu_P\bar{\sigma}_P}{(\sigma_P)^2+(\bar{\sigma}_P)^2};\\
\kappa^P=\frac{(\bar{\sigma}_P)^2}{(\sigma_P)^2};\ \ \
&\bar{\kappa}^P=\frac{(\bar{\sigma}_P)^2}{(\sigma_P)^2+(\bar{\sigma}_P)^2}.
\end{align*}


Our interest will be in pricing and hedging $\lambda$ units of a
contingent claim written on the non-traded assets $\mathcal{S}$ with
maturity $T$. The payoff is delivered at either a random time $\tau$
or maturity $T$. The random time $\tau$ represents an inter-temporal
default time, which is constructed in a canonical way. Let $e$ be an
independent exponential random variable on the same probability
space $(\Omega,\mathcal{F},\mathbf{P})$. Then the random time $\tau$
is constructed as follows
$$
\tau=\inf\left\{s\geq 0: \int_{0}^sa(\mathcal{S}_u,u)du\geq
e\right\},
$$
where $a(\cdot,\cdot)$ is an intensity function valued in
$\mathbb{R}_+$. The original Brownian filtration
$\mathbb{F}=\{\mathcal{F}_t\}_{t\geq 0}$ is enlarged by
$\mathcal{G}_t=\mathcal{F}_t\vee\mathcal{H}_t$ for $t\geq 0$ with
$\mathcal{H}_t=\sigma(\{\tau\leq u\}:0\leq u\leq t)$. Hence $\tau$
is the first arrival time of a Cox process (\emph{or doubly
stochastic Poisson process}), which satisfies the following
enlargement of filtration property: The stochastic process
$\mathcal{W}$ is an $(n+2)$-dimensional Brownian motion under both
filtrations $\{\mathcal{F}_t\}_{t\geq 0}$ and
$\{\mathcal{G}_t\}_{t\geq 0}$ (see Chapter 8 of Bielecki and
Rutkowski \cite{Bielecki-text} for details).

If such a random time $\tau$ happens before maturity $T$, i.e.
$\tau\leq T$, the payoff at $\tau$ is $R
\mathfrak{C}^{\lambda}(\mathcal{S}_{\tau},\tau)$, where $R\in[0,1]$
represents the recovery rate, and $\mathfrak{C}^{\lambda}:
\mathbb{R}^n\times[0,T]\rightarrow\mathbb{R}_+$ is the solution of
the PDE (\ref{PDE for n nontraded assets}). In Theorem
\ref{Markov_lemma_111} we shall show that
$\mathfrak{C}^{\lambda}(\cdot,\cdot)$ is actually the utility
indifference value function of the contingent claim, so we consider
the fractional recovery of market value.
If $\tau>T$, the payoff at $T$ is $\lambda g({\cal{S}}_{T})$, where
$g(\cdot)$ is a payoff function valued in $\mathbb{R}_+$. Therefore,
the total payoff for $\lambda$ units of this contingent claim is
$$\mathbf{1}_{\{\tau\leq T\}}R\mathfrak{C}^{\lambda}(\mathcal{S}_{\tau},\tau)
+\mathbf{1}_{\{\tau> T\}}\lambda g(\mathcal{S}_{T}).$$

\begin{assumption}
(i) The coefficients $\mu_i,\sigma_i,\bar{\sigma}_i$, the intensity
function $a$, and the payoff function $g$ depend on the logarithm
price of the non-traded assets:
$f=f(\mathcal{S}_t,t)\in\mathcal{C}_b^{1}(\mathbb{R}^n\times[0,T])$
for $f=\mu,\sigma_i,\bar{\sigma}_i,a,g$, where
$\mathcal{C}_b^{1}(\mathbb{R}^n\times[0,T])$ is the space of bounded
and continuous functions which are Lipschitz continuous in space.

(ii) The volatilities of non-traded assets $\mathcal{S}$ satisfy the
following uniformly elliptic condition: $|\sigma_i|\geq\epsilon>0$
for $i=1,\dots,n$.

\end{assumption}


The model is applicable in many situations. It might be that these
assets are (i) not traded at all, or (ii) that they are traded
illiquidly, or (iii) that they are in fact liquidly traded but the
investor concerned is not permitted to trade them for some reason.
Our main application is to the counterparty risk of derivatives
where the final payoff depends upon both the value of the
counterparty's assets and the assets underlying the derivative
itself (see Section 3). A second potential area of application is to
residual or basis risks arising when the assets used for hedging
differ from the assets underlying the contract in question (see
Davis \cite{Davis}). Typically this arises when the assets
underlying the derivative are illiquidly traded (case (ii) above)
and standardized futures contracts are used instead. Contracts may
involve several assets, for example, a spread option with payoff $
(K-e^{S^1_T}-e^{S^2_T})^+ $ or a basket option with payoff
$(K-e^{S^1_T}- \cdots- e^{S^n_T})^+$. Such contracts frequently
arise in applications to commodity, energy, and weather derivatives.
Finally, a one dimensional example of the situation in (iii) is that
of employee stock options (see \cite{Henderson-QF}).

On the other hand, the model also allows for intertemporal default
risk, and the recovery at the prepayment time is mark-to-market
value.
Hence, the model is well suited to counterparty credit risk, where
the intertemporal payment due to default usually depends on the
market value of the contract. Such an intertemporal default is
modeled in a reduced form, so the corresponding counterparty credit
risk model is a hybrid between the structural and reduced form
approaches. Other applications include optimal investment problems
with uncertain time horizon (see Blanchet-Scalliet et al
\cite{Blanchet}), and modeling the prepayment risk of
mortgage-backed securities (see Zhou \cite{ZhouT}).

Our approach is to consider the utility indifference valuation for
such a contingent claim. For this we need to consider the
optimization problem for the investor both with and without the
option. The investor has initial wealth $x\in \mathbb{R}$ at any
time $t\in[0,T]$, and is able to trade the financial index with
price $P_t$ (and riskless bond with price $1$). This will enable the
investor to partially hedge the risks she is exposed to via her
position in the claim. Depending on the context, the financial index
may be a stock, commodity or currency index, for example.







The holder of the option has an exponential utility function with
respect to her terminal wealth:
$$U_T(x)=-e^{-\gamma x}\ \ \ \text{for}\ \gamma\geq 0.$$

At time $t\in[0,T]$, the investor holds $\lambda$ units of the
contingent claim, whose price is denoted as
$\mathfrak{C}_t^{\lambda}$ and is to be determined, and invests her
remaining wealth $x-\mathfrak{C}_t^{\lambda}$ in the financial index
$P$. The investor will follow an admissible trading strategy:
\begin{equation*}
\pi\in\mathcal{A}_{\mathcal{F}}[t,T]=\left\{(\pi_s)_{s\in[t,T]}:\pi_s=\pi(\mathcal{S}_s,s)\
\text{is}\ \text{uniformly bounded}\right\},
\end{equation*}
which results in the wealth on the event $\{\tau>t\}$:
\begin{equation}\label{X_equ}
X_s^{x-\mathfrak{C}_t^{\lambda}}(\pi)=x-
\mathfrak{C}_t^{\lambda}+\int_t^s\frac{\pi_{s}}{P_s}dP_s.
\end{equation}

The investor will optimize over such strategies to choose an optimal
${\pi}^{*,\lambda}$ by maximizing her expected terminal utility:
\begin{equation}\label{optm1}
\esssup_{\pi\in\mathcal{A}_{\mathcal{F}}[t,T]}E^{\mathbf{P}}\left[-e^{-\gamma\left(\mathbf{1}_{\{
t<\tau\leq T\}}\left(X_{\tau}^{x-
\mathfrak{C}_t^{\lambda}}(\pi)+R\mathfrak{C}^{\lambda}(\mathcal{S}_{\tau},\tau)\right)+\mathbf{1}_{\{\tau>
T\}}\left(X_T^{x- \mathfrak{C}_t^{\lambda}}(\pi)+\lambda
g(\mathcal{S}_{T})\right)
 \right)}|\mathcal{G}_t\right].
\end{equation}

To define the utility indifference price for the option, we also
need to consider the optimization problem for the investor without
the option. Her wealth equation is the same as (\ref{X_equ}) but
starts from initial wealth $x$ and she will choose an optimal
$\pi^{*,0}$ by maximizing
\begin{equation}\label{optm2}
\esssup_{\pi\in\mathcal{A}_{\mathcal{F}}[t,T]}E^{\mathbf{P}}\left[-e^{-\gamma
X_T^{x}(\pi)}|\mathcal{G}_t\right].
\end{equation}
Since the payoff of the optimal portfolio problem (\ref{optm2}) is
$\mathbb{F}$-adapted, (\ref{optm2}) is equivalent to the standard
Merton problem on the event $\{\tau>t\}$:
$$\esssup_{\pi\in\mathcal{A}_{\mathcal{F}}[t,T]}E^{\mathbf{P}}\left[-e^{-\gamma
X_T^{x}(\pi)}|\mathcal{F}_t\right],$$ where the filtration is
restricted to $\{\mathcal{F}_t\}_{t\geq 0}$.

The utility indifference price for the option is the cash amount
that the investor is willing to pay such that she is no worse off in
expected utility terms than she would have been without the option.
For a general overview of utility indifference pricing, we refer to
the monograph edited by Carmona \cite{MR2547456} and
the survey article by Henderson and Hobson
\cite{Henderson3} therein.

\begin{definition}\label{definition1}

The utility indifference price $\mathfrak{C}_t^{\lambda}$ of
$\lambda$ units of the derivative with the payoff
$$\mathbf{1}_{\{\tau\leq
T\}}R\mathfrak{C}^{\lambda}(\mathcal{S}_{\tau},\tau)+\mathbf{1}_{\{\tau>
T\}}\lambda g(\mathcal{S}_{T})$$ at time $t\in[0,T]$ is defined by
the solution to
\begin{align}\label{def}
&\
\esssup_{\pi\in\mathcal{A}_{\mathcal{F}}[t,T]}E^{\mathbf{P}}\left[-e^{-\gamma\left(\mathbf{1}_{\{
t<\tau\leq T\}}\left(X_{\tau}^{x-
\mathfrak{C}_t^{\lambda}}(\pi)+R\mathfrak{C}^{\lambda}(\mathcal{S}_{\tau},\tau)\right)+\mathbf{1}_{\{\tau>
T\}}\left(X_T^{x- \mathfrak{C}_t^{\lambda}}(\pi)+\lambda
g(\mathcal{S}_{T})\right)
 \right)}|\mathcal{G}_t\right]\nonumber\\
=&\
\esssup_{\pi\in\mathcal{A}_{\mathcal{F}}[t,T]}E^{\mathbf{P}}\left[-e^{-\gamma
X_T^{x}(\pi)}|\mathcal{G}_t\right].
\end{align}
The hedging strategy for $\lambda$ units of the derivative at time
$t$ on the event $\{\tau>t\}$ is defined by the difference in the
optimal trading strategies ${\pi}^{*,\lambda}_t-\pi^{*,0}_t$.
\end{definition}

%


\subsection{Utility Indifference Price}

Our main result in this subsection is to show that the utility
indifference price is given by $\mathfrak{C}^{\lambda}_t
=\mathbf{1}_{\{\tau>t\}}\mathfrak{C}^{\lambda}(\mathcal{S}_t,t),$
where $\mathfrak{C}^{\lambda}(\mathcal{S}_t,t)$ is the solution of
the reaction-diffusion PDE (\ref{PDE for n nontraded assets}) with
quadratic gradients, so it can be interpreted as the (pre-default)
utility indifference value function. The function
$\mathfrak{C}^{\lambda}(\mathcal{S}_t,t)$ is the main object that we
are working on.
Define the following operators:
\begin{align}
\mathbf{L}^1&=\frac12\sum_{i=1}^n\sigma_i^2(\mathbf{s},t)\partial_{s_is_i}+\frac12\sum_{i,j=1}^n\bar{\sigma}_i
(\mathbf{s},t)\bar{\sigma}_j(\mathbf{s},t)\partial_{s_is_j}
+\sum_{i=1}^n\mu_i(\mathbf{s},t)\partial_{s_i},\\
\mathbf{L}^2&=-\sum_{i=1}^n\bar{\vartheta}^P\bar{\sigma}_i(\mathbf{s},t)\partial_{s_i}-\frac{\gamma}{2}\sum_{i=1}^n\sigma_i^2(\mathbf{s},t)(\partial_{s_i})^2
-\frac{\gamma}{2}\sum_{i,j=1}^n(1-\bar{\kappa}^P)\bar{\sigma}_i(\mathbf{s},t)\bar{\sigma}_j(\mathbf{s},t)(\partial_{s_i})(\partial_{s_j}),
\end{align}
and for $\mathbf{s}=(s_1,\cdots,s_n)$ and
$\mathfrak{C}^{\lambda}(\mathbf{s},t)$,
\begin{equation}
\mathbf{L}^3\mathfrak{C}^{\lambda}(\mathbf{s},t)
=\frac{a(\mathbf{s},t)}{\gamma}\left[1-e^{\gamma(1-R)\mathfrak{C}^{\lambda}(\mathbf{s},t)}\right].
\end{equation}
The operator $\mathbf{L}^1$ describes the infinitesimal behavior of
the price processes of the non-traded assets
$\mathcal{S}=(S^1,\cdots,S^n)$, the operator $\mathbf{L}^2$ reflects
the investor's risk aversion, and the operator $\mathbf{L}^3$
reflects the intertemporal payment which is distorted by the
investor's risk aversion.

\begin{theorem}\label{Markov_lemma_111} (PDE representation for utility indifference price)

Suppose that Assumption 2.1 is satisfied. Then the following
reaction-diffusion PDE with quadratic gradients on the domain
$(\mathbf{s},t)\in\mathbb{R}^n\times[0,T]$:
\begin{eqnarray}\label{PDE for n nontraded assets}
    \left\{
    \begin{array}{ll}
    -\partial_t\mathfrak{C}^{\lambda}(\mathbf{s},t)-(\mathbf{L}^1+
    \mathbf{L}^2+\mathbf{L}^3)\mathfrak{C}^{\lambda}(\mathbf{s},t)=0,
    \\
    \mathfrak{C}^{\lambda}(\mathbf{s},T)=\lambda g(\mathbf{s})
    \end{array}\right.
\end{eqnarray}
admits a unique viscosity solution
$\mathfrak{C}^{\lambda}(\mathbf{s},t)\in\mathcal{C}^{1}_b(\mathbb{R}^n\times[0,T])$.
Moreover, the utility indifference price of $\lambda$ units of the
derivative at time $t\in[0,T]$ with the payoff
$\mathbf{1}_{\{\tau\leq
T\}}R\mathfrak{C}^{\lambda}(\mathcal{S}_{\tau},\tau)+\mathbf{1}_{\{\tau>
T\}}\lambda g(\mathcal{S}_{T})$  is given by
$$\mathfrak{C}_t^{\lambda}=\mathbf{1}_{\{\tau> t\}}\mathfrak{C}^{\lambda}(\mathcal{S}_t,t),$$
and the hedging strategy for $\lambda$ units of the option at time
$t$ on the event $\{\tau> t\}$ is given by
\begin{equation}\label{hedge2}
-\frac{\bar{\kappa}^P}{\bar{\sigma}_P}\sum_{i=1}^n\bar{\sigma}_i(\mathcal{S}_t,t)\partial_{s_i}\mathfrak{C}^{\lambda}(\mathcal{S}_t,t).
\end{equation}
\end{theorem}

The proof of Theorem \ref{Markov_lemma_111} is provided in Appendix
\ref{appendix}, where we do not rely on the dynamic programming
principle. Instead, we transform the optimal portfolio problems
(\ref{optm1}) and (\ref{optm2}) into risk-sensitive control
problems, and derive a quadratic BSDE representation for the utility
indifference price, inspired by Theorem 2.3 of Henderson and Liang
\cite{HL}. Then PDE (\ref{PDE for n nontraded assets}) is obtained
by an application of nonlinear Feynman-Kac formula for quadratic
BSDE (see Kobylanski \cite{Kobylanski}). We note that the number of
units $\lambda$ only appears in the terminal condition. In the
following, we present the case $\lambda=1$, and the price at time
$t\in[0,T]$ is simply denoted by $\mathfrak{C}_t=\mathbf{1}_{\{\tau>
t\}}\mathfrak{C}(\mathcal{S}_t,t)$.

We first compare to the situation that the market is complete. If
the underlying assets $\mathcal{S}=(S^1,\cdots,S^n)$ could be
traded, the market would become complete, and the pricing and
hedging of the contingent claim with payoff:
$$\mathbf{1}_{\{\tau\leq
T\}}R\bar{\mathfrak{C}}(\mathcal{S}_{\tau},\tau)+\mathbf{1}_{\{\tau>
T\}} g(\mathcal{S}_{T})$$ falls into the multidimensional
Black-Scholes framework with intertemporal default risk.

\begin{corollary} \label{complete_lemma}
Suppose that Assumption 2.1 is satisfied, and that
$\mathcal{S}=(S^1,\cdots,S^n)$ are traded assets. Let
$\bar{\mathfrak{C}}(\mathbf{s},t)\in\mathcal{C}^{1}_b(\mathbb{R}^n\times[0,T])$
be the unique viscosity solution of the reaction-diffusion PDE on
the domain $(\mathbf{s},t)\in\mathbb{R}^n\times[0,T]$:
\begin{eqnarray}\label{JohnsonPDE}
    \left\{
    \begin{array}{ll}
    -\displaystyle\partial_t\bar{\mathfrak{C}}(\mathbf{s},t)
    -(\bar{\mathbf{L}}^1+\bar{\mathbf{L}}^3)\bar{\mathfrak{C}}(\mathbf{s},t)
    =0,\\
    \bar{\mathfrak{C}}(\mathbf{s},T)=g(\mathbf{s})
    \end{array}\right.
\end{eqnarray}
where the operators $\bar{\mathbf{L}}^1$ and $\bar{\mathbf{L}}^3$
are given, respectively, by
\begin{align*}
\bar{\mathbf{L}}^1&=\frac12\sum_{i=1}^n\sigma_i^2(\mathbf{s},t)\partial_{s_is_i}+
\frac12\sum_{i,j=1}^n\bar{\sigma}_i(\mathbf{s},t)\bar{\sigma}_j(\mathbf{s},t)\partial_{s_is_j}\\
\bar{\mathbf{L}}^3\bar{\mathfrak{C}}(\mathbf{s},t)&=-a(\mathbf{s},t)(1-R)\bar{\mathfrak{C}}(\mathbf{s},t).
\end{align*}
Then the price of the option with payoff $\mathbf{1}_{\{\tau\leq
T\}}R\bar{\mathfrak{C}}(\mathcal{S}_{\tau},\tau)+\mathbf{1}_{\{\tau>
T\}} g(\mathcal{S}_{T})$ at time $t\in[0,T]$ is given by
$\bar{\mathfrak{C}}_t=\mathbf{1}_{\{\tau>t\}}\bar{\mathfrak{C}}(\mathcal{S}_t,t)
$.

\end{corollary}

The pricing equation (\ref{PDE for n nontraded assets}) has an
additional nonlinear term $\mathbf{L}^2$ relative to the complete
market pricing PDE (\ref{JohnsonPDE}), and this $\mathbf{L}^2$
reflects the investor's risk aversion. Moreover, the inter-temporal
payment term $\bar{\mathbf{L}}^3$ in (\ref{JohnsonPDE}) is distorted
to $\mathbf{L}^3$ by the investor's risk aversion in (\ref{PDE for n
nontraded assets}). We have the following asymptotic result relating
the utility indifference price $\mathfrak{C}_t$ to the complete
market price $\bar{\mathfrak{C}}_t$ at any time $t\in[0,T]$.

\begin{proposition} \label{Asy2} Assume that
\begin{equation} \label{CAPM}
 \bar{\vartheta}^P=\frac{\mu_i}{\bar{\sigma}_i}(\mathbf{s},t)\ \text{for}\ i=1,\dots,n.
\end{equation}
Then the unit utility indifference value function
$\mathfrak{C}(\mathbf{s},t)$ uniformly converges to the complete
market value function $\bar{\mathfrak{C}}(\mathbf{s},t)$ as
$\gamma\rightarrow 0$ on any compact subset of
$\mathbb{R}^n\times[0,T]$.

\end{proposition}

\begin{proof} By the condition (\ref{CAPM}), the first-order linear
terms in (\ref{PDE for n nontraded assets}) become zero:
$$\sum_{i=1}^n\mu_i(\mathbf{s},t)\partial_{s_i}-\sum_{i=1}^n\bar{\vartheta}^P\bar{\sigma}_i(\mathbf{s},t)\partial_{s_i}=0.$$
When $\gamma\rightarrow 0$, the terms involving $\gamma$ in
$\mathbf{L}^2$ converge to zero, and
$\mathbf{L}^3\mathfrak{C}(\mathbf{s},t)\rightarrow\bar{\mathbf{L}}^3\mathfrak{C}(\mathbf{s},t)$.
Therefore, by the stability of viscosity solutions, there exists a
subsequence $\gamma_n\rightarrow 0$ such that the viscosity
solutions of (\ref{PDE for n nontraded assets}), denoted by
$\mathfrak{C}(\mathbf{s},t;\gamma_n)$, uniformly converge to
$\bar{\mathfrak{C}}(\mathbf{s},t)$ on any compact subset of
$\mathbb{R}^n\times[0,T]$, where $\bar{\mathfrak{C}}(\mathbf{s},t)$
satisfies $(\ref{JohnsonPDE})$.
\end{proof}

The restriction (\ref{CAPM}) in fact corresponds to a relation
between the Sharpe ratios of the non-traded assets $\mathcal{S}$ and
the financial index $P$. The Sharpe ratio of $P$ is given by
$\sqrt{\bar{\theta}^P}$. Similarly, we define the Sharpe ratio of
$S^i$ to be $\sqrt{\bar{\theta}^i}$, where $\bar{\theta}^i =
\frac{\mu_i^2}{\sigma_i^2+\bar{\sigma}_i^2}(\mathbf{s},t)$. Then
(\ref{CAPM}) is equivalent to the relation;
$$\sqrt{\bar{\theta}^i} =
\left(\frac{\bar{\sigma}_i(\mathbf{s},t) \bar{\sigma}_P }{\sqrt{
\bar{\sigma}_i^2(\mathbf{s},t) + \sigma_i^2(\mathbf{s},t) } \sqrt{
\bar{\sigma}_P^2 + \sigma_P^2} }\right)\sqrt{\bar{\theta}^P} =
\rho_{i P} \sqrt{\bar{\theta}^P},
$$ where $\rho_{i P}$ is the correlation between $S^i$ and $P$. This
corresponds to the relation we expect from the capital asset pricing
model (CAPM) when assets are traded. Since not all assets are traded
here, we would not necessarily expect (\ref{CAPM}) to hold. The
intuition is that when the idiosyncratic volatilities disappear, and
when assets are traded, there cannot be a difference in using the
financial index $P$ or the assets themselves to hedge.

Based on the pricing equation (\ref{PDE for n nontraded assets}) and
the PDE comparison principle, we present a number of monotone
properties of the utility indifference price. Their proofs are
similar to Section \ref{sec-boundedsolution}, so we omit them.


\begin{proposition}\label{Asy1}
The unit utility indifference value function
$\mathfrak{C}(\mathbf{s},t)$ is increasing with the recovery rate
$R$, the payoff $g(\cdot)$ and the intensity $a(\cdot,\cdot)$, and
is decreasing with the risk aversion parameter $\gamma$. Moreover,
if the condition (\ref{CAPM}) holds, then
$\mathfrak{C}(\mathbf{s},t)$ is also decreasing in the idiosyncratic
volatility of the traded asset $\sigma_P^2$ (or its proportion of
total volatility, $1-\bar{\kappa}^P$).
\end{proposition}

The last assertion of the above proposition tells us that the higher
the idiosyncratic volatility $\sigma_{P}^2$ of the traded asset
(\emph{or as a proportion of total volatility}), the worse it is as
a hedging instrument, and the lower the price one is willing to pay.
This generalizes the monotonicity obtained in the one dimensional
non-traded asset model (see, for example, Henderson
\cite{Henderson-QF} and Frei and Schweizer \cite{FSI} in a
non-Markovian model with stochastic correlation).

\subsection{A Semigroup Approximation by Splitting} 

For a reaction-diffusion PDE with quadratic gradients like (\ref{PDE
for n nontraded assets}), it is not possible to obtain an explicit
solution. A special case where an explicit solution does exist is
the one dimensional version without intertemporal default. Taking
$n=1$, $\sigma_{P}=0$ and $R=1$ in (\ref{PDE for n nontraded
assets}) recovers the pricing PDE of \cite{Henderson2},
\cite{Henderson1} and \cite{Musiela}, which is solved by the
Cole-Hopf transformation. However, this transformation does not
apply directly to our multidimensional problem (\ref{PDE for n
nontraded assets}) because the coefficients of the quadratic
gradient terms in $\mathbf{L}^2$ do not match, and the existence of
the intertemporal payment term $\mathbf{L}^3$.
Instead, we will develop a splitting algorithm which will enable us
to take advantage again of the Cole-Hopf transformation to linearize
the PDEs.


The splitting method (\emph{or fractional step, prediction and
correction, etc}) can be dated back to Marchuk \cite{Marchuk} in the
late 1960's. The application of splitting to nonlinear PDEs such as
HJB equations is difficult mainly because of the verification of the
convergence for the approximate scheme. This was overcome by Barles
and Souganidis \cite{Barles}, who employed the idea of viscosity
solutions and proved that any \emph{monotone, stable} and
\emph{consistent} numerical scheme converges provided there exists a
comparison principle for the limiting equation.


The idea of splitting in our setting is the following. The  time
derivative of the pricing PDE (\ref{PDE for n nontraded assets})
depends on the sum of semigroup operators (\emph{or the associated
infinitesimal operators}) corresponding to the different factors.
These semigroups usually are of different nature. For each
subproblem corresponding to each semigroup there might be an
effective way providing solutions. For the sum of these semigroups,
however, we usually can not find an accurate method. Hence,
application of splitting method means that instead of the sum, we
treat the semigroup operators separately.

The tricky part is how to split the equation (\emph{or how to group
factors}) effectively. In next lemma, we separate the pricing PDE
(\ref{PDE for n nontraded assets}) into three pricing factors by
using the transformation (\ref{logoperator}), which is the key step
to apply the splitting method to (\ref{PDE for n nontraded assets}).

\begin{lemma} Define a new
differential operator:
\begin{equation} \label{logoperator}
\frac{\partial}{\partial \eta}=
\sum_{i=1}^n\bar{\sigma}_i(\mathbf{s},t)\frac{\partial}{\partial
s_i}.
\end{equation}
Then (\ref{PDE for n nontraded assets}) reduces to
\begin{equation}\label{PDE101}
-\partial_t\mathfrak{C}-\left(\hat{\mathbf{L}}^1+\hat{\mathbf{L}}^2+\hat{\mathbf{L}}^3\right)\mathfrak{C}=0,
\end{equation}
where
\begin{align}
\hat{\mathbf{L}}^{1}&=\frac12\partial_{\eta\eta}-\frac{\gamma}{2}(1-\bar{\kappa}^P)(\partial_{\eta})^2,\\
\hat{\mathbf{L}}^{2}&=\frac12\sum_{i=1}^n\sigma_i^2(\mathbf{s},t)\partial_{s_is_i}+\sum_{i=1}^nA_i(\mathbf{s},t)\partial_{s_i}
-\frac{\gamma}{2}\sum_{i=1}^n\sigma_i^2(\mathbf{s},t)(\partial_{s_i})^2\\
\hat{\mathbf{L}}^3&=\mathbf{L}^3
\end{align}
with
\begin{equation*} \label{Ai}
A_i(\mathbf{s},t)=\mu_i(\mathbf{s},t)-\frac{1}{2}\left[\sigma_i^2(\mathbf{s},t)+\bar{\sigma}_i^2(\mathbf{s},t)\right]-\bar{\vartheta}^P\bar{\sigma}_i(\mathbf{s},t).
\end{equation*}
\end{lemma}


For any $0\leq t<t+\Delta \leq T$ and any function
$\phi\in\mathcal{C}^1_b(\mathbb{R}^{n+1})$, we define the following
nonlinear backward semigroup operators $\mathbf{S}^i(\Delta)$ by
$\phi(\cdot)\mapsto \mathfrak{C}^i(\cdot,t)$ where
\begin{equation}\label{semigroupequation}
-\partial_t\mathfrak{C}^i-\hat{\mathbf{L}}^i \mathfrak{C}^i=0;\ \ \
\ \mathfrak{C}^i(\cdot,t+\Delta)=\phi(\cdot)
\end{equation}
on the domain $\mathbb{R}^{n+1}\times[t,t+\Delta]$ for $i=1,2, 3$.
That is, $\mathbf{S}^i(\Delta)\phi$ is the solution of
(\ref{semigroupequation}) at time $t$ with terminal data $\phi$ at
time $t+\Delta$.

We observe that (\ref{semigroupequation}) for $i=1, 2$ can be
linearized by Cole-Hopf transformations. Indeed, by letting
$\bar{\mathfrak{C}}^1=\exp(-\gamma(1-\bar{\kappa}^P)
\mathfrak{C}^1)$, then we have $\bar{\mathfrak{C}}^1$ satisfying
\begin{equation} \label{C1}
-\partial_t\bar{\mathfrak{C}}^1-\frac12\partial_{\eta\eta}\bar{\mathfrak{C}}^1=0.
\end{equation}
By letting $\bar{\mathfrak{C}}^2=\exp(-\gamma \mathfrak{C}^2)$, then
we have $\bar{\mathfrak{C}}^2$ satisfying
\begin{equation} \label{C2}
-\partial_t\bar{\mathfrak{C}}^2-\frac12\sum_{i=1}^n\sigma_i^2(\mathbf{s},t)\partial_{s_is_i}\bar{\mathfrak{C}}^2-\sum_{i=1}^nA_i(\mathbf{s},t)\partial_{s_i}\bar{\mathfrak{C}}^2=0.
\end{equation}
Moreover, (\ref{semigroupequation}) for $i=3$ can be approximated by
Picard iterations, since $\hat{\mathbf{L}}^3=\mathbf{L}^3$ is
Lipschitz continuous (see Appendix \ref{sec-boundedsolution}).

\begin{lemma}\label{semigroup}
The operators $\mathbf{S}^i(\Delta)$ for $i=1,2,3$ have the
following properties:
\begin{itemize}
\item (i) For any function $\phi\in\mathcal{C}^{1}_b(\mathbb{R}^{n+1})$,
$$\lim_{\Delta\downarrow 0}\mathbf{S}^i(\Delta)\phi=\phi$$
uniformly on any compact subset of $\mathbb{R}^{n+1}$.

\item (ii)
$$\mathbf{S}^i(\Delta')\phi=\mathbf{S}^i(\Delta'-\Delta)\mathbf{S}^i(\Delta)\phi$$
for any $0\leq t<t+\Delta<t+\Delta'\leq T$.

\item (iii) $$\mathbf{S}^i(0)\phi=\phi.$$
((i) (ii) and (iii) ensure that $\mathbf{S}^i(\Delta)$ is indeed a
strongly continuous semigroup operator.)

\item (iv) For any functions $\phi,\psi\in\mathcal{C}^{1}_b(\mathbb{R}^{n+1})$ such that $\phi\geq \psi$,
$$\mathbf{S}^i(\Delta)\phi\geq\mathbf{S}^i(\Delta)\psi.$$

\item (v) $\mathbf{S}^i(\Delta)\phi$ is uniformly bounded, and
moreover,
$$|\mathbf{S}^i(\Delta)\phi-\mathbf{S}^i(\Delta)\psi|_{0}\leq C|\phi-\psi|_{0},$$
where $|\cdot|_{0}$ represents the usual supremum norm.


\item (vi) For any $\phi\in\mathcal{C}_b^{\infty}(\mathbb{R}^{n+1})$, the space of
bounded and smooth functions, define the consistent error:
$$\mathcal{E}^i(\Delta,\phi)=\left|\frac{\mathbf{S}^i(\Delta)\phi(\mathbf{s})-\phi(\mathbf{s})}{\Delta}-\hat{\mathbf{L}}^i\phi(\mathbf{s})\right|_0.$$
Then
\begin{align*}
\mathcal{E}^1(\Delta,\phi)\leq &\ C\Delta
\left(|\partial_{\mathbf{s}}^4\phi|_{0}
+\sum_{i+j=4}|\partial_{\mathbf{s}}^i\phi|_{0}|\partial_{\mathbf{s}}^j\phi|_{0}
+|\partial_{\mathbf{s}}^2\phi|_{0}|\partial_{\mathbf{s}}\phi|^2_{0}\right)\\
&\
+C\Delta^2\left(|\partial_{\mathbf{s}}^3\phi|^2_0+|\partial_{\mathbf{s}}^2\phi|_0^2|\partial_{\mathbf{s}}\phi|^2_0\right),\\
\mathcal{E}^2(\Delta,\phi)\leq &\ C\Delta
\left(\sum_{i=2,3,4}|\partial_{\mathbf{s}}^i\phi|_{0}
+\sum_{i+j=3,4}|\partial_{\mathbf{s}}^i\phi|_{0}|\partial_{\mathbf{s}}^j\phi|_{0}
+|\partial_{\mathbf{s}}^2\phi|_{0}|\partial_{\mathbf{s}}\phi|^2_{0}\right)\\
&\
+C\Delta^2\left(\sum_{i=2,3}|\partial_{\mathbf{s}}^i\phi|^2_0+|\partial_{\mathbf{s}}^2\phi|_0^2|\partial_{\mathbf{s}}\phi|^2_0\right),\\
\mathcal{E}^3(\Delta,\phi)\leq &\ C\Delta,
\end{align*}
where $\mathbf{s}=(\eta,s_1,\dots,s_{n})$ with a slight abuse of
notation.
\end{itemize}
\end{lemma}

\begin{proof} (i)-(v) are immediate. We only prove (vi) in the
following. The key idea is to make use of the coupled FBSDE
representation for
$\mathfrak{C}^i(\cdot,t)=\mathbf{S}^i(\Delta)\phi(\cdot)$. We first
prove the case $i=1$.

Let $(\Omega,\mathcal{F},\mathbb{F}=\{\mathcal{F}_t\}_{t\geq
0},\mathbf{Q})$ be a filtered probability space satisfying the
\emph{usual conditions}, on which supports a one-dimensional
Brownian motion $B$. Then an application of nonlinear Feynman-Kac
formula yields the following FBSDE representation of
$\mathfrak{C}^1(\eta,t)$:
\begin{equation}\label{representation_1}
\mathfrak{C}^1(\eta,t)=\phi(X_{t+\Delta})-\int_t^{t+\Delta}\partial_{\eta}\mathfrak{C}^1(X_s,s)dB_s
={E}^{\mathbf{Q}}[\phi(X_{t+\Delta})|X_t=\eta],
\end{equation}
with
\begin{equation}\label{representation_11}
X_s=\eta-\int_t^{s}\frac{\gamma}{2}(1-\bar{\kappa}^{P})\partial_{\eta}\mathfrak{C}^1(X_u,u)du+\int_t^sdW_u.
\end{equation}

By applying It\^o's formula to $\phi(X_{t+\Delta})$, we obtain that
\begin{align*}
  &\ \mathbf{S}^1(\Delta)\phi(\eta) -\phi(\eta) - \Delta\hat{\mathbf{L}}^1\phi(\eta) \\
= &\
E^{\mathbf{Q}}[\phi(X_{t+\Delta})|X_t=\eta] - \phi(\eta) - \Delta\hat{\mathbf{L}}^1\phi(\eta)\\
= &\
E^{\mathbf{Q}}\left[\phi(\eta)+\int_t^{t+\Delta}\partial_{\eta}\phi(X_s)dX_s+\frac12\partial^2_{\eta}\phi(X_s)d\langle
X\rangle_s|X_t=\eta\right] - \phi(\eta) - \Delta\hat{\mathbf{L}}^1\phi(\eta)\\
=&\
E^{\mathbf{Q}}\left[\int_t^{t+\Delta}[-\frac{\gamma}{2}(1-\bar{\kappa}^P)\partial_{\eta}\phi(X_s)\partial_{\eta}\mathfrak{C}^1(X_s,s)+\frac12\partial^2_{\eta}\phi(X_s)]ds|X_t=\eta\right]-\Delta\hat{\mathbf{L}}^1\phi(\eta)
\end{align*}
Since $$\partial_{\eta}\mathfrak{C}^1(X_s,s) =
\partial_{\eta} E^{\mathbf{Q}}[\phi(X_{t+\Delta})|X_s] =
E^{\mathbf{Q}}[\partial_{\eta}\phi(X_{t+\Delta})Y_{t+\Delta}|X_s],$$
with $Y$ being the gradient flow of $X$:
$Y_{t+\Delta}=1-\int_s^{t+\Delta}\frac{\gamma}{2}(1-\bar{\kappa}^P)\partial_{\eta\eta}\mathfrak{C}^1(X_u,u)Y_udu$,
we further have
\begin{align*}
  &\ \mathbf{S}^1(\Delta)\phi(\eta) -\phi(\eta) - \Delta\hat{\mathbf{L}}^1\phi(\eta) \\
= &\
E^{\mathbf{Q}}\left[\int_t^{t+\Delta}-\frac{\gamma}{2}(1-\bar{\kappa}^P)\partial_{\eta}\phi(X_s)\partial_{\eta}\phi(X_{t+\Delta})Y_{t+\Delta}+\frac12\partial^2_{\eta}\phi(X_s)ds|X_t=\eta\right]-\Delta\hat{\mathbf{L}}^1\phi(\eta).
\end{align*}

By applying It\^o's formula once again, we calculate the above
conditional expectation as
\begin{align*}
E^{\mathbf{Q}}\left[-\frac{\gamma}{2}(1-\bar{\kappa}^P)\left(\Delta\partial_{\eta}\phi(\eta)
+\int_t^{t+\Delta}\int_t^s
(-\frac{\gamma}{2}(1-\bar{\kappa}^P)\sum_{i+j=3}\partial_{\eta}^i\phi\partial_{\eta}^j\mathfrak{C}^1
+\frac12\partial_{\eta}^3\phi) duds\right)\right.\\
\cdot\left(\partial_{\eta}\phi(\eta)+\int_t^{t+\Delta}(
-\frac{\gamma}{2}(1-\bar{\kappa}^P)\sum_{i+j=3}\partial_{\eta}^i\phi\partial_{\eta}^j\mathfrak{C}^1
+\frac12\partial_{\eta}^3\phi)ds \right)\cdot(1+{O}(\Delta))\\
\left.+\frac12\left( \Delta\partial_{\eta}^2\phi(\eta)
+\int_t^{t+\Delta}\int_t^s(-\frac{\gamma}{2}(1-\bar{\kappa}^P)
\sum_{i+j=4}\partial_{\eta}^i\phi\partial_{\eta}^{j}\mathfrak{C}^1+\frac12\partial_{\eta}^4\phi)
duds\right)|X_t=\eta\right]
\end{align*}
where $i,j\geq 1$ are integers, and we omit the arguments in
$\phi(\cdot)$ and $\mathfrak{C}^1(\cdot,\cdot)$. Since
$\phi\in\mathcal{C}_b^{\infty}$ and using the fact that
$\partial_{\eta}^i\mathfrak{C}^1(X_s,s)=E^{\mathbf{Q}}[\partial_{\eta}^i\phi(X_{t+\Delta})|X_s](1+{O}(\Delta))$,
we obtain the consistent error estimate:
\begin{align}\label{estimate1}
&\ \mathcal{E}^1(\Delta,\phi)\notag\\
\leq &\ C
\Delta\left(|\partial_{\eta}^4\phi|_0+|\partial_{\eta}^3\phi|_0|\partial_{\eta}\phi|_0+|\partial_{\eta}^2\phi|_0^2+|\partial_{\eta}^2\phi|_0|\partial_{\eta}\phi|_0^2\right)
+C\Delta^2\left(|\partial_{\eta}^3\phi|^2_0+|\partial_{\eta}^2\phi|_0^2|\partial_{\eta}\phi|^2_0\right).
\end{align}


The proof for the case $i=2$ is similar, so we skip its proof and
only provide the corresponding consistent error estimate:
\begin{align}\label{estimate2}
&\ \mathcal{E}^2(\Delta,\phi)\notag\\
\leq &\ C
\Delta\left(|\partial_{\mathbf{s}}^4\phi|_0+|\partial_{\mathbf{s}}^3\phi|_0|\partial_{\mathbf{s}}\phi|_0+|\partial_{\mathbf{s}}^2\phi|_0^2+|\partial_{\mathbf{s}}^2\phi|_0|\partial_{\mathbf{s}}\phi|_0^2
+|\partial_{\mathbf{s}}^3\phi|_0+|\partial_{\mathbf{s}}^2\phi|_0|\partial_{\mathbf{s}}\phi|_0+|\partial_{\mathbf{s}}^2\phi|_0\right)\notag\\
&\
+C\Delta^2\left(|\partial_{\mathbf{s}}^3\phi|^2_0+|\partial_{\mathbf{s}}^2\phi|_0^2|\partial_{\mathbf{s}}\phi|^2_0+|\partial_{\mathbf{s}}^2\phi|_0^2\right).
\end{align}

Finally, the proof for case $i=3$ is a simple application of Taylor
expansion, which yields
\begin{equation}\label{estimate3}
\mathcal{E}^3(\Delta,\phi)\leq C\Delta.
\end{equation}
\end{proof}

Next we use semigroup operators $\mathbf{S}^i(\Delta)$ for $i=1,2,3$
to give the semigroup approximation for the solution of PDE
(\ref{PDE101}) (\emph{or PDE (\ref{PDE for n nontraded assets}) with
different coordinates}), which is the main result of this section.

\begin{theorem} (Semigroup approximation for utility indifference price)
\label{semigrouptheorem}

Suppose that Assumption 2.1 is satisfied. Let $t_i=i\Delta$ for
$i=0,1,\dots,N$ with $N\Delta=T$. Then the unit utility indifference
value function $\mathfrak{C}(\cdot,\cdot)$ of the derivative with
the payoff $\mathbf{1}_{\{\tau\leq
T\}}R\mathfrak{C}(\mathcal{S}_{\tau},\tau)+\mathbf{1}_{\{\tau> T\}}
g(\mathcal{S}_{T})$ at time $t_i$ is approximated by
\begin{equation}\label{splitting}
\mathfrak{C}^{\Delta}(\cdot,t_i)=\mathbf{S}(\Delta)\mathfrak{C}^{\Delta}(\cdot,t_{i+1})
=\mathbf{S}^3(\Delta)\mathbf{S}^2(\Delta)\mathbf{S}^1(\Delta)\mathfrak{C}^{\Delta}(\cdot,t_{i+1})
\end{equation}
with $\mathfrak{C}^{\Delta}(\cdot,t_N)=g(\cdot)$. The values between
any two adjacent partition points are obtained by usual linear
interpolation. Then,
$$\lim_{\Delta\rightarrow 0}\mathfrak{C}^{\Delta}(\cdot,\cdot)=\mathfrak{C}(\cdot,\cdot).$$
uniformly on any compact subset of $\mathbb{R}^{n+1}\times[0,T]$.
\end{theorem}

\begin{proof} The proof is based on the
Barles-Souganidis monotone scheme \cite{Barles}, in which they
proved that any \emph{monotone, stable} and \emph{consistent}
numerical scheme converges, provided there exists a comparison
principle for the limiting equation.

By the stability property (v) of Lemma \ref{semigroup}, the
following semi-relaxed limits of $\mathfrak{C}^{\Delta}$ are well
defined:
$$\overline{\mathfrak{C}}(\mathbf{s},t)=\limsup_{(\mathbf{s}',t')\rightarrow(\mathbf{s},t),
\atop \Delta\rightarrow 0}\mathfrak{C}^{\Delta}(\mathbf{s}',t');\ \
\
\underline{\mathfrak{C}}(\mathbf{s},t)=\liminf_{(\mathbf{s}',t')\rightarrow(\mathbf{s},t),
\atop \Delta\rightarrow 0}\mathfrak{C}^{\Delta}(\mathbf{s}',t').
$$
We show that $\overline{\mathfrak{C}}$ is a viscosity subsolution of
(\ref{PDE101}). A symmetric argument will imply that
$\underline{\mathfrak{C}}$ is a viscosity supersolution of
(\ref{PDE101}), which proves that
$\overline{\mathfrak{C}}=\underline{\mathfrak{C}}=\mathfrak{C}$, so
$\mathfrak{C}^{\Delta}$ converges to $\mathfrak{C}$ locally
uniformly.

Let $\phi\in\mathcal{C}^{\infty}_{b}$ and $(\mathbf{s}_0,t_0)$ be
such that
$$0=(\overline{\mathfrak{C}}-\phi)(\mathbf{s}_0,t_0)=\max_{(\mathbf{s}',t')}
(\overline{\mathfrak{C}}-\phi)(\mathbf{s}',t').$$ By the definition
of $\overline{\mathfrak{C}}$, there exists a sequence
$(\mathbf{s}_n,t_n)$ such that
$$(\mathbf{s}_n,t_n,\Delta)\rightarrow(\mathbf{s}_0,t_0,0),\ \ \text{and}\ \ \mathfrak{C}^{\Delta}(\mathbf{s}_n,t_n)\rightarrow\overline{\mathfrak{C}}(\mathbf{s}_0,t_0).$$
Moreover, by extracting a subsequence if necessary,
$(\mathbf{s}_n,t_n)$ is also the maximum point of
$\mathfrak{C}^{\Delta}-\phi$:
$$\delta^{\Delta}=(\mathfrak{C}^{\Delta}-\phi)(\mathbf{s}_n,t_n)=\max_{(\mathbf{s}',t')}
(\mathfrak{C}^{\Delta}-\phi)(\mathbf{s}',t')\rightarrow 0.$$ The
monotone property (iv) of Lemma \ref{semigroup} then implies that
\begin{align*}
&\frac{\phi(\mathbf{s}_n,t_n)+\delta^{\Delta}-\mathbf{S}(\Delta)\left(\phi(\mathbf{s}_n,t_n+\Delta)+\delta^{\Delta}\right)}{\Delta}\\
\leq & \frac{\mathfrak{C}^{\Delta}(\mathbf{s}_n,t_n)-
\mathbf{S}(\Delta)\mathfrak{C}^{\Delta}(\mathbf{s}_n,t_n+\Delta)}{\Delta}=0
\end{align*}
From (v) of Lemma \ref{semigroup},
$\mathbf{S}(\Delta)\left(\phi(\mathbf{s}_n,t_n+\Delta)+\delta^{\Delta}\right)\leq
\mathbf{S}(\Delta)\left(\phi(\mathbf{s}_n,t_n+\Delta)\right)+C\delta^{\Delta}$.
In turn, using the consistent property (vi) of Lemma \ref{semigroup}
together with the estimate (\ref{estimate4}) and letting
$(\mathbf{s}_n,t_n,\Delta)\rightarrow(\mathbf{s}_0,t_0,0)$, we
obtain that
$$-\partial_t\phi(\mathbf{s}_0,t_0)-
(\hat{\mathbf{L}}^1+\hat{\mathbf{L}}^2+\hat{\mathbf{L}}^3)\phi(\mathbf{s}_0,t_0)\leq
0.$$ That is, $\overline{\mathfrak{C}}(\cdot,\cdot)$ is a viscosity
subsolution of (\ref{PDE101}).
\end{proof}

\subsection{The Convergence Rate of Splitting}

The proof of the convergence rate of numerical viscosity solutions
is much more difficult than the proof of the convergence itself.
This open problem was first solved by Krylov \cite{Krylov}, who
developed the so called \emph{shaking the coefficients technique},
and was extensively studied by Barles and Jakobsen in a series of
their papers, which also introduced an alternative \emph{optimal
switching approximation} method (see \cite{BJ} and \cite{Jakobsen}
with more references therein).

The key step to apply Krylov's technique in our setting is to obtain
an estimate of consistent error, and the other steps follow a
similar line of argument as in \cite{BJ} and \cite{Jakobsen}. Hence,
we only derive the consistent error estimate in the following, and
outline the proofs of the other steps.

We first derive an upper bound of
$\mathfrak{C}-\mathfrak{C}^{\Delta}$. For any $\epsilon>0$, we may
build a viscosity subsolution
$\mathfrak{C}^{\epsilon}\in\mathcal{C}^{1}_b(\mathbb{R}^{n+1}\times[0,T])$
of (\ref{PDE101}) by shaking its coefficients such that
\begin{equation}\label{upperbound_1}
\mathfrak{C}-\epsilon\leq \mathfrak{C}^{\epsilon}\leq \mathfrak{C}.
\end{equation}

Next, we regularize $\mathfrak{C}^{\epsilon}$ by convolution to
obtain a smooth subsolution of (\ref{PDE101}). For this, let $\rho$
be an $\mathbb{R}_+$-valued smooth function with support
$\{|\mathbf{s}|< 1\}\times\{0<t<1\}$ and mass $1$. From this
function, we introduce a sequence of mollifiers $\rho_{\epsilon}$ as
follows:
$$\rho_{\epsilon}(\mathbf{s},t)=\frac{1}{\epsilon^{n+3}}\rho\left(\frac{\mathbf{s}}{\epsilon},\frac{t}{\epsilon^2}\right).$$
Define
$$\mathfrak{C}_{\epsilon}(\mathbf{s},t)=\mathfrak{C}^{\epsilon}*
\rho_{\epsilon}(\mathbf{s},t)=\int_{-\epsilon^2< \tau< 0}\int_{|e|<
\epsilon}\mathfrak{C}^{\epsilon}(\mathbf{s}-e,t-\tau)\rho_{\epsilon}(e,\tau)ded\tau.$$

A Riemann sum approximation shows that
$\mathfrak{C}_{\epsilon}(\mathbf{s},t)$ can be viewed as the limit
of convex combinations of
$\mathfrak{C}^{\epsilon}(\mathbf{s}-e,t-\tau)$ for $-\epsilon^2\leq
\tau\leq 0$ and $|e|<\epsilon$. Since (\ref{PDE101}) is convex in
$\partial_t\mathfrak{C}$ and $\partial_{\mathbf{s}}^i\mathfrak{C}$
for $i=0,1,2$, the convex combinations of
$\mathfrak{C}^{\epsilon}(\mathbf{s}-e,t-\tau)$ are also the
subsolutions of (\ref{PDE101}). In turn, $\mathfrak{C}_{\epsilon}$
is itself a subsolution of (\ref{PDE101}) by the stability of
viscosity solutions.

On the other hand, standard properties of mollifiers imply that
$\mathfrak{C}_{\epsilon}\in\mathcal{C}_b^{\infty}$,
\begin{equation}\label{upperbound_2}
|\mathfrak{C}^{\epsilon}-\mathfrak{C}_{\epsilon}|_0\leq C\epsilon,
\end{equation}
and moreover,
\begin{equation}\label{mollifier}
|\partial_{t}^i\mathfrak{C}_{\epsilon}|_0\leq C\epsilon^{1-2i};\ \ \
|\partial_{\mathbf{s}}^j\mathfrak{C}_{\epsilon}|_0\leq
C\epsilon^{1-j}.
\end{equation}

The key result to obtain the convergence rate of splitting is the
following consistent error estimate.

\begin{lemma} Suppose that Assumption 2.1 is satisfied.
Then,
\begin{align}\label{estimate4}
\mathcal{E}(\Delta,\mathfrak{C}_{\epsilon})=&\
\left|\frac{\mathbf{S}(\Delta)\mathfrak{C}_{\epsilon}(\mathbf{s},t)-\mathfrak{C}_{\epsilon}(\mathbf{s},t-\Delta)}{\Delta}-\partial_t\mathfrak{C}_{\epsilon}(\mathbf{s},t)-
(\hat{\mathbf{L}}^1+\hat{\mathbf{L}}^2+\hat{\mathbf{L}}^3)\mathfrak{C}_{\epsilon}(\mathbf{s},t)\right|_0\notag\\
\leq&\
C\left(\Delta+\Delta\epsilon^{-3}+\Delta^2\epsilon^{-5}+\Delta^3\epsilon^{-6}+\Delta^4\epsilon^{-8}+\Delta^6\epsilon^{-10}\right).
\end{align}
\end{lemma}

\begin{proof}
Define
$\mathfrak{C}_{\epsilon}^i=\mathfrak{C}_{\epsilon}^{i-1}+\Delta\hat{\mathbf{L}}^i\mathfrak{C}_{\epsilon}^{i-1}$
for $i=1,2,3$ with
$\mathfrak{C}_{\epsilon}^0=\mathfrak{C}_{\epsilon}$. We then split
the consistent error
$\Delta\mathcal{E}(\Delta,\mathfrak{C}_{\epsilon})$ into five parts
as follows:
\begin{align*}
&|\mathbf{S}(\Delta)\mathfrak{C}_{\epsilon}(\mathbf{s},t)-\mathfrak{C}_{\epsilon}(\mathbf{s},t-\Delta)-\Delta\partial_t\mathfrak{C}_{\epsilon}(\mathbf{s},t)-\Delta(\hat{\mathbf{L}}^1+\hat{\mathbf{L}}^2+\hat{\mathbf{L}}^3)\mathfrak{C}_{\epsilon}(\mathbf{s},t)|_0\\
=&\
|\mathbf{S}(\Delta)\mathfrak{C}_{\epsilon}(\mathbf{s},t)-\mathfrak{C}_{\epsilon}(\mathbf{s},t)-\Delta(\hat{\mathbf{L}}^1+\hat{\mathbf{L}}^2+\hat{\mathbf{L}}^3)\mathfrak{C}_{\epsilon}(\mathbf{s},t)\\
&+
\mathfrak{C}_{\epsilon}(\mathbf{s},t)-\mathfrak{C}_{\epsilon}(\mathbf{s},t-\Delta)-\Delta\partial_t\mathfrak{C}_{\epsilon}(\mathbf{s},t)|_0\\
\leq&\
|\mathbf{S}^3(\Delta)\mathbf{S}^2(\Delta)\mathbf{S}^1(\Delta)\mathfrak{C}_{\epsilon}(\mathbf{s},t)-\mathbf{S}^3(\Delta)\mathbf{S}^2(\Delta)\mathfrak{C}^1_{\epsilon}(\mathbf{s},t)|_0\\
&+
|\mathbf{S}^3(\Delta)\mathbf{S}^2(\Delta)\mathfrak{C}^1_{\epsilon}(\mathbf{s},t)-\mathbf{S}^3(\Delta)\mathfrak{C}^2_{\epsilon}(\mathbf{s},t)|_0\\
&+
|\mathbf{S}^3(\Delta)\mathfrak{C}_{\epsilon}^2(\mathbf{s},t)-\mathfrak{C}^3_{\epsilon}(\mathbf{s},t)|_0\\
&+
|\mathfrak{C}_{\epsilon}^3(\mathbf{s},t)-\mathfrak{C}_{\epsilon}(\mathbf{s},t)-\Delta(\hat{\mathbf{L}}^1+\hat{\mathbf{L}}^2+\hat{\mathbf{L}}^3)\mathfrak{C}_{\epsilon}(\mathbf{s},t)|_0\\
&+
|\mathfrak{C}_{\epsilon}(\mathbf{s},t)-\mathfrak{C}_{\epsilon}(\mathbf{s},t-\Delta)-\Delta\partial_t\mathfrak{C}_{\epsilon}(\mathbf{s},t)|_0\\
=&\ [1]+[2]+[3]+[4]+[5].
\end{align*}

For $[1]$, using (v) of Lemma \ref{semigroup} and the consistent
error estimate (\ref{estimate1}), we have
\begin{align}\label{estimate5}
&\
|\mathbf{S}^3(\Delta)\mathbf{S}^2(\Delta)\mathbf{S}^1(\Delta)\mathfrak{C}_{\epsilon}(\mathbf{s},t)-\mathbf{S}^3(\Delta)\mathbf{S}^2(\Delta)\mathfrak{C}^1_{\epsilon}(\mathbf{s},t)|_0\notag\\
\leq &\
C|\mathbf{S}^1(\Delta)\mathfrak{C}_{\epsilon}(\mathbf{s},t)-\mathfrak{C}^1_{\epsilon}(\mathbf{s},t)|_0
\leq C\Delta(\Delta\epsilon^{-3}+\Delta^2\epsilon^{-4}),
\end{align}
where the second inequality follows from the property
(\ref{mollifier}) of the mollifier $\mathfrak{C}_{\epsilon}$, while
keeping the worst (\emph{highest order}) terms involving $\epsilon$.

For $[2]$, we employ (v) of Lemma \ref{semigroup}, the consistent
error estimate (\ref{estimate2}), together with the property
(\ref{mollifier}) of the mollifier $\mathfrak{C}^1_{\epsilon}$ to
derive that
\begin{equation}\label{estimate6}
|\mathbf{S}^3(\Delta)\mathbf{S}^2(\Delta)\mathfrak{C}^1_{\epsilon}(\mathbf{s},t)-\mathbf{S}^3(\Delta)\mathfrak{C}^2_{\epsilon}(\mathbf{s},t)|_0
\leq C\Delta
(\Delta\epsilon^{-3}+\Delta^2\epsilon^{-5}+\Delta^3\epsilon^{-6}+\Delta^4\epsilon^{-8}+\Delta^6\epsilon^{-10}),
\end{equation}
and similarly for $[3]$,
\begin{equation}\label{estimate7}
|\mathbf{S}^3(\Delta)\mathfrak{C}_{\epsilon}^2(\mathbf{s},t)-\mathfrak{C}^3_{\epsilon}(\mathbf{s},t)|_0\leq
C\Delta^2.
\end{equation}

Next, for $[4]$, by the definition of $\mathfrak{C}^i_{\epsilon}$
with $i=1,2,3$, we have
\begin{align}\label{estimate8}
&\
|\mathfrak{C}_{\epsilon}^3(\mathbf{s},t)-\mathfrak{C}_{\epsilon}(\mathbf{s},t)-\Delta(\hat{\mathbf{L}}^1+\hat{\mathbf{L}}^2+\hat{\mathbf{L}}^3)\mathfrak{C}_{\epsilon}(\mathbf{s},t)|_0\notag\\
=&\
\Delta\left|(\hat{\mathbf{L}}^2\mathfrak{C}^1_{\epsilon}(\mathbf{s},t)-\hat{\mathbf{L}}^2\mathfrak{C}_{\epsilon}(\mathbf{s},t))
+(\hat{\mathbf{L}}^3\mathfrak{C}^2_{\epsilon}(\mathbf{s},t)-\hat{\mathbf{L}}^3\mathfrak{C}_{\epsilon}(\mathbf{s},t))\right|_0\notag\\
\leq&\
C\Delta\left((\Delta\epsilon^{-3}+\Delta^2\epsilon^{-4})+(\Delta\epsilon^{-1}+\Delta^2\epsilon^{-3}+\Delta^3\epsilon^{-4})\right)\notag\\
\leq&\
C\Delta(\Delta\epsilon^{-3}+\Delta^2\epsilon^{-4}+\Delta^3\epsilon^{-4}).
\end{align}

For $[5]$, using (v) of Lemma \ref{semigroup} and Taylor expansion,
we get
\begin{align}\label{estimate9}
&\
|\mathfrak{C}_{\epsilon}(\mathbf{s},t)-\mathfrak{C}_{\epsilon}(\mathbf{s},t-\Delta)-\Delta\partial_t\mathfrak{C}_{\epsilon}(\mathbf{s},t)|_0\notag\\
\leq&\
|\int_{t-\Delta}^{t}\left(\partial_{t}\mathfrak{C}_{\epsilon}(\mathbf{s},t)-\int_v^{t}\partial_t^2\mathfrak{C}_{\epsilon}(\mathbf{s},u)du\right)dv-
\Delta\partial_t\mathfrak{C}_{\epsilon}(\mathbf{s},t)|_0\notag\\
\leq&\ C\Delta^2|\partial_t^{2}\mathfrak{C}_{\epsilon}|_0\leq
C\Delta^2\epsilon^{-3}.
\end{align}

Finally, the consistent error estimate (\ref{estimate4}) follows
from (\ref{estimate5})-(\ref{estimate9}) by keeping the worst terms
involving $\epsilon$.
\end{proof}

\begin{theorem}\label{theorem_error_1}
(Upper bound of convergence rate)

Suppose that Assumption 2.1 is satisfied. Then the difference
between the unit utility indifference value function
$\mathfrak{C}(\cdot,\cdot)$ and its approximation
$\mathfrak{C}^{\Delta}(\cdot,\cdot)$ has an upper bound:
$$\mathfrak{C}-\mathfrak{C}^{\Delta}\leq C\Delta^{\frac14}.$$
\end{theorem}

\begin{proof}
Using the fact that $\mathfrak{C}_{\epsilon}$ is a smooth
subsolution of (\ref{PDE101}) together with the consistent error
estimate (\ref{estimate4}), we obtain that
\begin{equation*}
\frac{\mathfrak{C}_{\epsilon}(\mathbf{s},t-\Delta)-\mathbf{S}(\Delta)\mathfrak{C}_{\epsilon}(\mathbf{s},t)}{\Delta}\leq
C\left(\Delta+\Delta\epsilon^{-3}+\Delta^2\epsilon^{-5}+\Delta^3\epsilon^{-6}+\Delta^4\epsilon^{-8}+\Delta^6\epsilon^{-10}\right).
\end{equation*}
On the other hand,
$\mathfrak{C}^{\Delta}(\mathbf{s},t-\Delta)=\mathbf{S}(\Delta)\mathfrak{C}^{\Delta}(\mathbf{s},t)$.
Hence,
\begin{align*}
\mathfrak{C}_{\epsilon}(\mathbf{s},t-\Delta)-\mathfrak{C}^{\Delta}(\mathbf{s},t-\Delta)\leq&\
\mathbf{S}(\Delta)\mathfrak{C}_{\epsilon}(\mathbf{s},t)-\mathbf{S}(\Delta)\mathfrak{C}^{\Delta}(\mathbf{s},t)\\
&+C\left(\Delta^2+\Delta^2\epsilon^{-3}+\Delta^3\epsilon^{-5}+\Delta^4\epsilon^{-6}+\Delta^5\epsilon^{-8}+\Delta^7\epsilon^{-10}\right)\\
\leq&\
C|\mathfrak{C}_{\epsilon}(\mathbf{s},t)-\mathfrak{C}^{\Delta}(\mathbf{s},t)|_0\\
&+C\left(\Delta^2+\Delta^2\epsilon^{-3}+\Delta^3\epsilon^{-5}+\Delta^4\epsilon^{-6}+\Delta^5\epsilon^{-8}+\Delta^7\epsilon^{-10}\right),
\end{align*}
where we used (v) of Lemma (\ref{semigroup}) in the second
inequality. We repeat the above procedure $T/\Delta$ times and get
\begin{equation}\label{upperbound_3}
\mathfrak{C}_{\epsilon}-\mathfrak{C}^{\Delta}\leq
C\left(\Delta+\Delta\epsilon^{-3}+\Delta^2\epsilon^{-5}+\Delta^3\epsilon^{-6}+\Delta^4\epsilon^{-8}+\Delta^6\epsilon^{-10}\right).
\end{equation}

In turn, (\ref{upperbound_1}), (\ref{upperbound_2}) and
(\ref{upperbound_3}) imply that
\begin{align*}
\mathfrak{C}-\mathfrak{C}^{\Delta}=&\
(\mathfrak{C}-\mathfrak{C}^{\epsilon})+
(\mathfrak{C}^{\epsilon}-\mathfrak{C}_{\epsilon})+(\mathfrak{C}_{\epsilon}-\mathfrak{C}^{\Delta})\\
\leq&\
C\left(\epsilon+\Delta+\Delta\epsilon^{-3}+\Delta^2\epsilon^{-5}+\Delta^3\epsilon^{-6}+\Delta^4\epsilon^{-8}+\Delta^6\epsilon^{-10}\right)\leq
C\Delta^{\frac14}
\end{align*}
by choosing $\epsilon=\Delta^{\frac14}$.
\end{proof}

To get a lower bound of $\mathfrak{C}-\mathfrak{C}^{\Delta}$, we
interchange the role of (\ref{PDE101}) and its splitting algorithm
(\ref{splitting}). For some technical reasons, we only consider the
full recovery rate $R=1$ when deriving the lower bound. Note that in
this situation $\mathbf{S}^3(\Delta)$ is an identity operator.
First, we need to rewrite the splitting algorithm (\ref{splitting})
as the following equation so that the \emph{elliptic condition}
holds:
\begin{equation}\label{scheme}
\frac{\mathfrak{C}^{\Delta}(\mathbf{s},t-\Delta)-\mathbf{S}(\Delta)\mathfrak{C}^{\Delta}(\mathbf{s},t)}{\Delta}=0.
\end{equation}

We shake the coefficients of (\ref{scheme}) to construct its
subsolution $\mathfrak{C}^{\Delta,\epsilon}$, and regularize it by
convolution
$\mathfrak{C}^{\Delta}_{\epsilon}=\mathfrak{C}^{\Delta,\epsilon}*\rho_{\epsilon}$.
Since
$\mathbf{S}(\Delta)\phi=\mathbf{S}^1(\Delta)\mathbf{S}^2(\Delta)\phi$
is concave in $\phi$, Jensen's inequality implies that
\begin{align*}
&\frac{\mathfrak{C}^{\Delta}_{\epsilon}(\mathbf{s},t-\Delta)-\mathbf{S}(\Delta)\mathfrak{C}^{\Delta}_{\epsilon}(\mathbf{s},t)}{\Delta}\notag\\
\leq&\ \int_{-\epsilon^2< \tau< 0}\int_{|e|<
\epsilon}\frac{\mathfrak{C}^{\Delta,\epsilon}(\mathbf{s}-e,t-\Delta-\tau)-\mathbf{S}(\Delta)\mathfrak{C}^{\Delta,\epsilon}(\mathbf{s}-e,t-\tau)}{\Delta}\rho_{\epsilon}(e,\tau)ded\tau\leq
0.
\end{align*}
Together with the consistent error estimate (\ref{estimate4}) with
$\mathfrak{C}_{\epsilon}$ replaced by
$\mathfrak{C}_{\epsilon}^{\Delta}$, we derive that
$$-\partial_t\mathfrak{C}_{\epsilon}^{\Delta}-(\hat{\mathbf{L}}^1+\hat{\mathbf{L}}^2)\mathfrak{C}_{\epsilon}^{\Delta}\leq C\left(\Delta+\Delta\epsilon^{-3}+\Delta^2\epsilon^{-5}+\Delta^3\epsilon^{-6}+\Delta^4\epsilon^{-8}+\Delta^6\epsilon^{-10}\right).$$
Thus,
$\mathfrak{C}_{\epsilon}^{\Delta}-(T-t)C\left(\Delta+\Delta\epsilon^{-3}+\Delta^2\epsilon^{-5}+\Delta^3\epsilon^{-6}+\Delta^4\epsilon^{-8}+\Delta^6\epsilon^{-10}\right)$
is a subsolution of (\ref{PDE101}). In turn, the comparison
principle of (\ref{PDE101}) implies that
$$\mathfrak{C}_{\epsilon}^{\Delta}-\mathfrak{C}\leq(T-t)C\left(\Delta+\Delta\epsilon^{-3}+\Delta^2\epsilon^{-5}+\Delta^3\epsilon^{-6}+\Delta^4\epsilon^{-8}+\Delta^6\epsilon^{-10}\right).$$
Combining with
$|\mathfrak{C}^{\Delta}-\mathfrak{C}_{\epsilon}^{\Delta}|_0\leq
C\epsilon$, we obtain the lower bound of the splitting algorithm
(\ref{splitting}):
\begin{theorem}\label{theorem_error_2}
(Lower bound of convergence rate)

Suppose that Assumption 2.1 is satisfied, and that the recovery
rate $R=1$. Then the difference between the unit utility
indifference value function $\mathfrak{C}(\cdot,\cdot)$ and its
approximation $\mathfrak{C}^{\Delta}(\cdot,\cdot)$ has a lower
bound:
$$\mathfrak{C}-\mathfrak{C}^{\Delta}\geq -C\Delta^{\frac14}.$$
\end{theorem}

\section{Application to Counterparty Risk}
\label{sec-application}

\subsection{Utility Indifference Valuation of Vulnerable Options}

In this section, we apply our multidimensional non-traded assets
model to consider the counterparty risk of derivatives with possible
default at maturity. Our concern as the buyer or holder of the
option is that the writer or counterparty may default on the option
with payoff $h(S_{T})$ at maturity $T$ and we will not receive the
full payoff. We have in mind several examples. A natural example is
that of a commodity producer who is writing options as part of a
hedging program (e.g. collars). Some of these options may be written
on illiquidly traded assets and thus the option holder is subject to
basis risk and in addition, is concerned with the default risk of
the option writer. A second example is the default risk of a
financial institution who has sold options on various underlying
assets - stocks, foreign exchange or commodities. In addition to the
possibility of basis risk, the buyer of these options does not
always have the ability to trade the underlying asset, or perhaps
they choose not to (they may be using the derivative as part of a
hedge already). A further example may be that of a purchaser of
insurance concerned with the default risk of the insurer. Typically
the insured party does not trade at all, which motivates our
consideration of this special case. Finally, the option holder may
be an employee of a company who receives employee stock options if
the company remains solvent. She is restricted from trading the
stock of the company, but can trade other indices or stocks in the
market. In contrast to the other examples, here the assets of the
counterparty and the underlying stock are those of the same company.

Consider an option written on an underlying asset with logarithm
price $S^1$ with payoff $h(S^1_T)$ at maturity $T$. Counterparty
default is modeled by comparing the value of the counterparty's
assets $\exp(S^2)$ to a default threshold $D$ at maturity, which
depends on the liabilities of the counterparty. Following Klein
\cite{Klein} we consider the situation $D=L$, where $L$ refers to
the option writer's liabilities, assumed to be a constant.
Generalizations to $D=f(S^1_T)$ are easily incorporated and allow
for the option liability itself to influence default, e.g. $f(x) =
h(x) + L$ was considered by Klein and Inglis \cite{KI2001} in a risk
neutral setting. If the writer defaults, the holder will receive the
proportion $h(S^1_T)/L$ of the assets $\exp(S^2)$ that her option
represents of the writer's liabilities, scaled to reflect a
proportional deadweight loss of $\alpha \in [0,1]$. The payoff of
the {\it vulnerable option} taking counterparty default into account
is
\begin{equation}\label{payoff_1}
g(S^1_T,S^2_T) = h(S^1_T) \emph{1}_{\{\exp(S^2_T)\geq
L\}}+(1-\alpha)\frac{h(S^1_T)}{L}
\exp(S^2_T)\emph{1}_{\{\exp(S^2_T)<L\}}.
\end{equation}
Note that $g\in\mathcal{C}^1_b(\mathbb{R}^2)$ except at a singular
point $S_T^2=\ln L$, if the payoff
$h\in\mathcal{C}^1_{b}(\mathbb{R})$, for example, a put option with
$(K-e^{S_T^1})^+$. To apply the splitting algorithm, we first need
to approximate $g$ by a sequence of (nondecreasing) Lipschitz
continuous and bounded functions $g^{\epsilon}$. For the numerical
simulation purpose, we only need to choose one $g^{\epsilon}$ for
$\epsilon$ small enough.

Note that the above payoff (\ref{payoff_1}) of the vulnerable option
written on $S^1$ taking account of the counterparty default can also
be regarded as the payoff of a basket option written on $(S^1,S^2)$
without taking account of intertemporal default risk, so it falls
into the framework of Section \ref{sec-model}. The underlying asset
$S^1$ and the counterparty's assets $S^2$ are both taken to be
non-traded assets so $n=2$ and their logarithm prices follow
(\ref{dynamics of nontraded assets_2}). The option holder faces some
unhedgeable price risk (due to $S^1$) and some unhedgeable
counterparty default risk (due to $S^2$). She can partially hedge
risks by trading the financial index $P$ following (\ref{P_equ_2}).

Our focus in the implementation on default at maturity enables us to
compare our results to several benchmark models - Johnson and Stulz
\cite{Johnson}, Klein \cite{Klein} and Klein and Inglis
\cite{KI2001} (see also \cite{LR2007} for intertemporal default).
Each takes a structural approach to price vulnerable options in a
complete market setting, and obtains two dimensional Black-Scholes
style formulas.
Implicit in this prior literature are the twin assumptions that the
asset underlying the option and the assets of the counterparty can
be traded, and therefore, can be used to hedge the counterparty risk
of derivatives.

Our use of the utility indifference approach is motivated by its
recent use in credit risk modeling where the concern is the default
of the reference name rather than the default of the counterparty.
Utility based pricing has also been utilized by Bielecki and
Jeanblanc \cite{Bielecki1}, Sircar and Zariphopolou \cite{MR2642963}
and recently Jiao et al \cite{Jiao1} \cite{Jiao2} in an intensity
based setting. Several authors have applied it in modeling of
defaultable bonds where the problem remains one dimensional, see in
particular Leung et al \cite{leung}, Jaimungal and Sigloch
\cite{JS2009}, and Liang and Jiang \cite{LJ2009}. In contrast,
options subject to counterparty risk are a natural situation where
two or more dimensions arise.

Based on Theorem \ref{semigrouptheorem}, we give the following
approximation scheme for the unit utility indifference price
$\mathfrak{C}(s_1,s_2,0)$ of the vulnerable option. Following
(\ref{logoperator}) we define a new operator:
$$\frac{\partial}{\partial\eta}=\bar{\sigma}_1\frac{\partial}{\partial
s_1}+\bar{\sigma}_2\frac{\partial}{\partial s_2}.$$

\begin{itemize}
  \item (i) Partition $[0,T]$ into $N$ equal intervals:
  $$0=t_0<t_1<\cdots<t_N=T.$$
  \item (ii) On $[t_{N-1},t_N]$, predict the solution by solving the
  following PDE with the given terminal data $g^{\epsilon}$:
 \begin{eqnarray*}
    \left\{
    \begin{array}{ll}
    \partial_t\mathfrak{C}^1+\frac12\partial_{\eta\eta}\mathfrak{C}^1-\frac{\gamma}{2}(1-\bar{\kappa}^P)(\partial_{\eta}\mathfrak{C}^1)^2=0,\\[+0.4cm]
    \mathfrak{C}^1|_{t=t_{N}}=g^{\epsilon}.&
    \end{array}\right.
\end{eqnarray*}
The above equation can be linearized via the Cole-Hopf
transformation:
$$\bar{\mathfrak{C}}^1=\exp(-\gamma(1-\bar{\kappa}^P)\mathfrak{C}^1).$$
Thus, we obtain $\mathfrak{C}^1|_{t=t_{N-1}}$ by solving the
corresponding linear PDE.

  \item (iii) On $[t_{N-1},t_N]$, correct the solution by solving the
  following PDE with the terminal data
  $\mathfrak{C}^1|_{t=t_{N-1}}$:
\begin{eqnarray*}
    \left\{
    \begin{array}{ll}
    \partial_{t}\mathfrak{C}^2+\frac12\sigma_1^2\partial_{s_1s_1}\mathfrak{C}^2+\frac12\sigma_2^2\partial_{s_2s_2}\mathfrak{C}^2+A_1\partial_{s_1}\mathfrak{C}^2+
A_2\partial_{s_2}\mathfrak{C}^2\\[+0.4cm]
\ \ \ \ \ \ \ -\frac{\gamma}{2}\sigma_1^2(\partial_{s_1}\mathfrak{C}^2)^2-\frac{\gamma}{2}\sigma_2^2(\partial_{s_2}\mathfrak{C}^2)^2=0,\\[+0.4cm]
    \mathfrak{C}^2|_{t=t_{N}}=\mathfrak{C}^1|_{t=t_{N-1}}
    \end{array}\right.
\end{eqnarray*}
where $A_1, A_2$ are given in (\ref{Ai}) to be
$A_i=\mu_i-\frac{1}{2}(\sigma_i^2+\bar{\sigma}_i^2)-\bar{\vartheta}^P\bar{\sigma}_i;
\: \: i=1,2$. The above equation can also be linearized by making
the exponential transformation: $$\bar{\mathfrak{C}}^2=\exp(-\gamma
\mathfrak{C}^2).$$ Thus, we obtain $\mathfrak{C}^2|_{t=t_{N-1}}$ by
solving the corresponding linear PDE, which is used as the
approximation of $\mathfrak{C}|_{t=t_{N-1}}$.

  \item (iv) Repeat the above procedure on $[t_{N-2},t_{N-1}]$, and
  obtain $\mathfrak{C}|_{t=t_{N-2}}$ $\dots$.
\end{itemize}

\subsection{Numerical Results}

We present results for the  European put with payoff $h(S^1_T) =
\left(K-e^{S^1_T}\right)^+$. If $S^1$ and $S^2$ are positively
correlated, this means when the put option is valuable
(in-the-money), the firm's assets $S^2$ tend to be small, so there
is a high risk of default. It is important to take counterparty risk
into account for puts in this case, as it will have a relatively
large impact on the price. (This would be even more significant when
the default trigger involves the option liability). However, for a
call, when the call is in-the-money, there is little default risk,
and so counterparty risk is less important. Unless otherwise stated,
the parameters are: $K=150$; $T=1$; $\exp(S^1)=50$; $\exp(S^2)=100$;
$L=1000$; $\alpha=0.05$; $\gamma = 1$; $\mu_P = 0.1$; $\sigma_P =
0.15$; $\bar{\sigma}_P = 0.2$; $\mu_1 = 0.15$; $\sigma_1 = 0.25$;
$\bar{\sigma}_1 = 0.3$; $\mu_2 = 0.1$; $\sigma_2 = 0.3$;
$\bar{\sigma}_2 =0.2$. These parameters result in correlation
between the underlying asset and firm's assets of $\rho_{1 2}=0.4 $;
and correlations between each asset and the financial index $P$ of $
\rho_{1 P}=0.6;\ \ \rho_{2 P}=0.4.$.

\begin{figure} 
\begin{minipage}[t]{0.5\linewidth}
\centering
\includegraphics[scale=0.3]{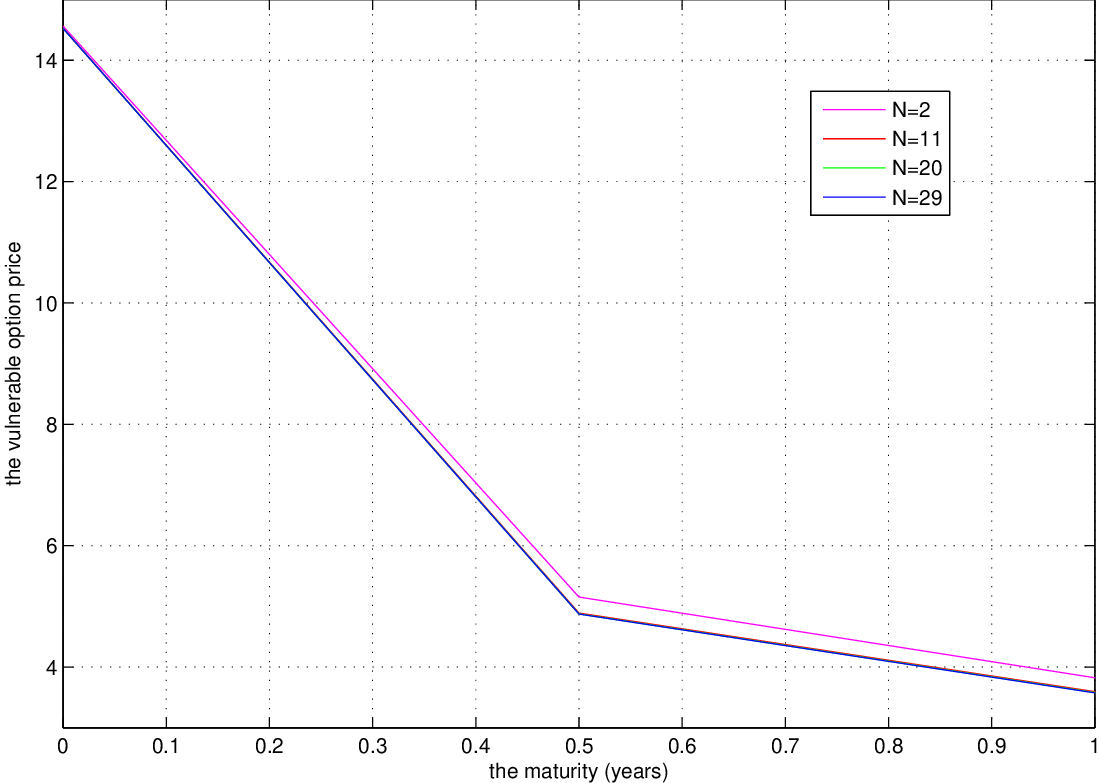}
\end{minipage}
\begin{minipage}[t]{0.5\linewidth}
\centering
\includegraphics[scale=0.3]{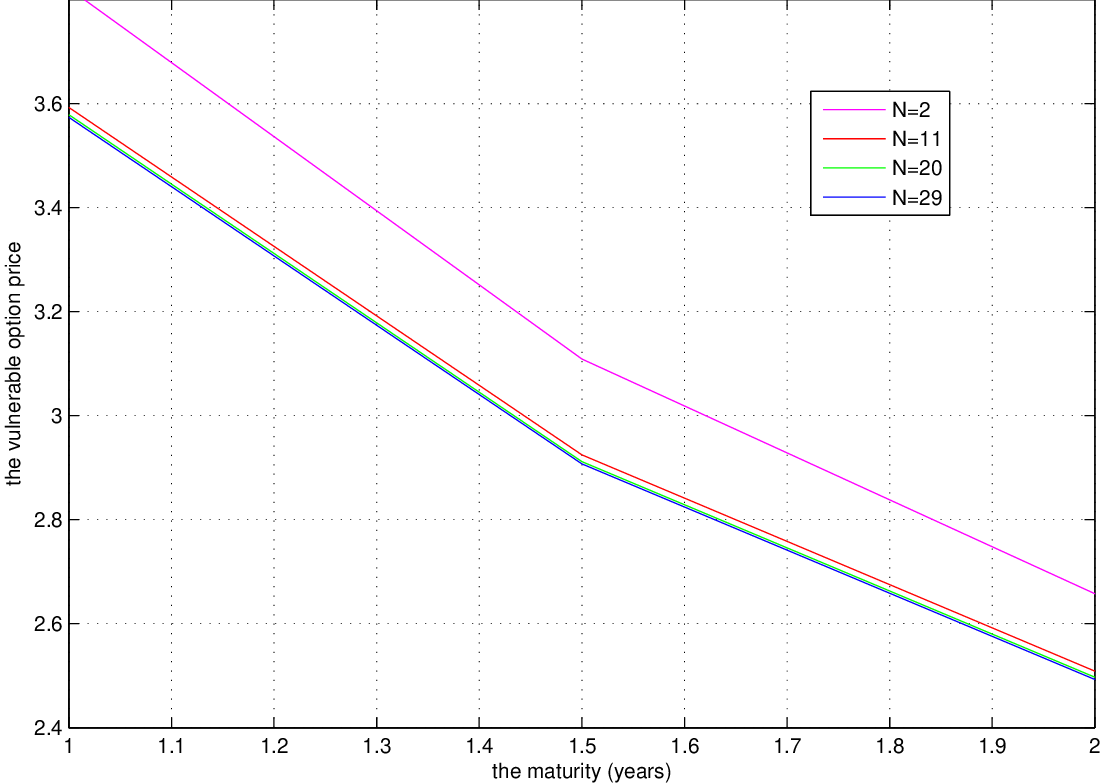}
\end{minipage}
\caption{Approximation of the option price for various time steps
$N$. The left panel takes $T\in[0,1]$; the right panel takes
$T\in[1,2]$.} \label{Fig-N}
\end{figure}

\begin{figure} 
\centering
\includegraphics[scale=0.35]{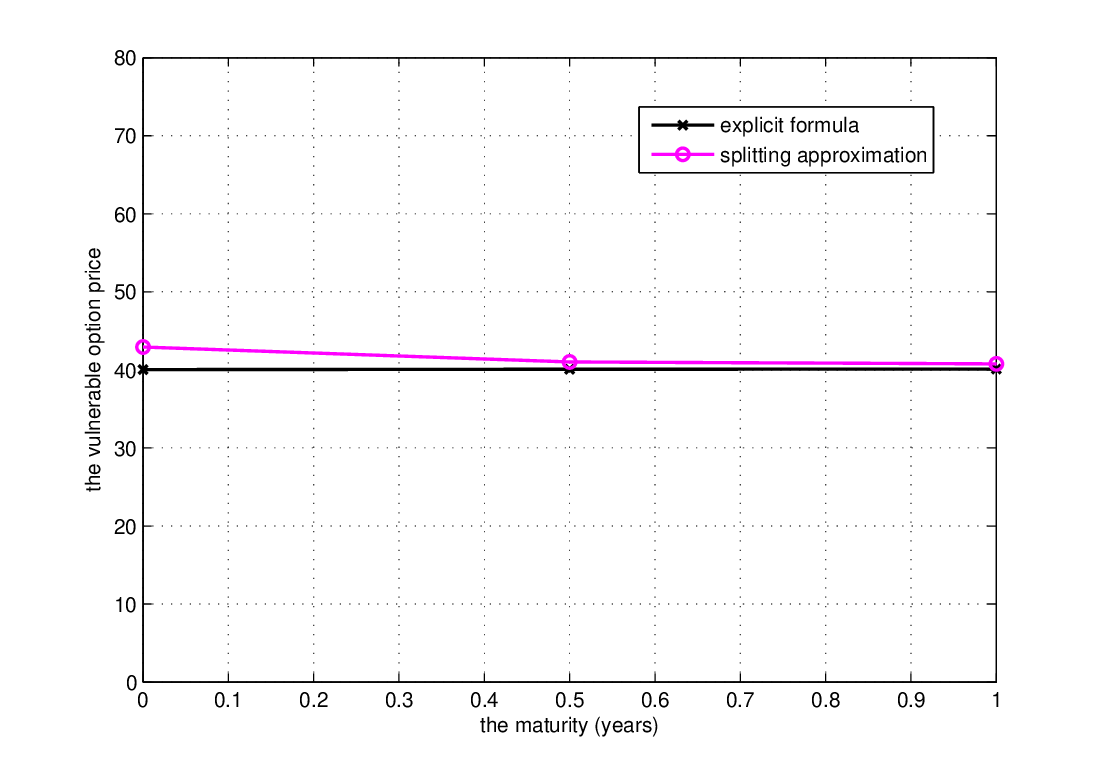}
\caption{Comparison of the explicit price and the approximated price
via splitting when $\bar{\sigma}^P=0$,
$\exp(S^2)=1400$, and $N=11$.}\label{Fig-approx}
\end{figure}

In Figure \ref{Fig-N} we show how the approximation converges as we
increase the number of time steps $N$. For our parameter values,
$N=11$ steps is sufficient for the prices to converge and we use it
in all subsequent figures. We aim to compare the utility
indifference price with hedging in the {\it financial index} with
the benchmark risk neutral price in a complete market (computed as
in Corollary \ref{complete_lemma} with $n=2$, as studied in Klein
\cite{Klein}). We also compare to the situation where the financial
index is independent of the other assets and thus there is {\it no
hedging} carried out. This price has an explicit formula:
$$-\frac{1}{\gamma}\ln E^{\mathbf{P}}
\left[-e^{-\gamma g^{\epsilon}(S_T^1,S_T^2)}\right].$$ Figure
\ref{Fig-approx} provides a demonstration of the accuracy of the
algorithm. We take $\bar{\sigma}_{P} = 0$ and compare the splitting
approximation to the above formula.

\begin{figure}
\begin{minipage}[t]{0.5\linewidth}
\centering
\includegraphics[scale=0.35]{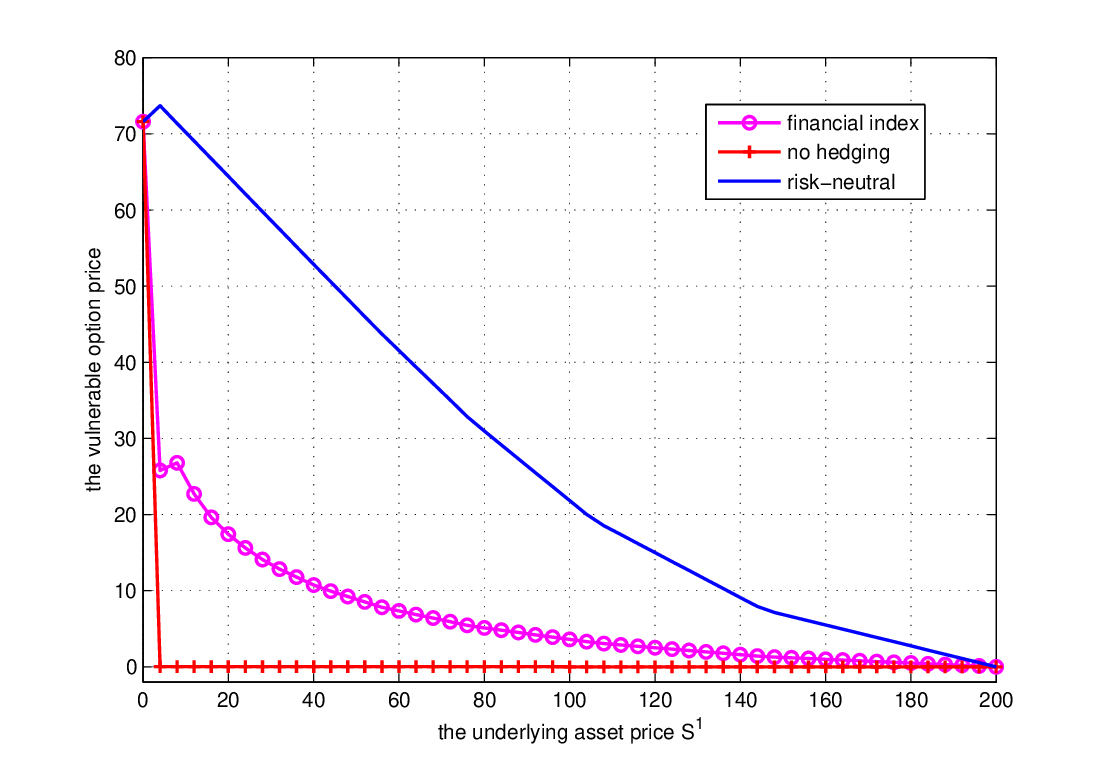}
\label{Fig1}
\end{minipage}
\begin{minipage}[t]{0.5\linewidth}
\centering
\includegraphics[scale=0.35]{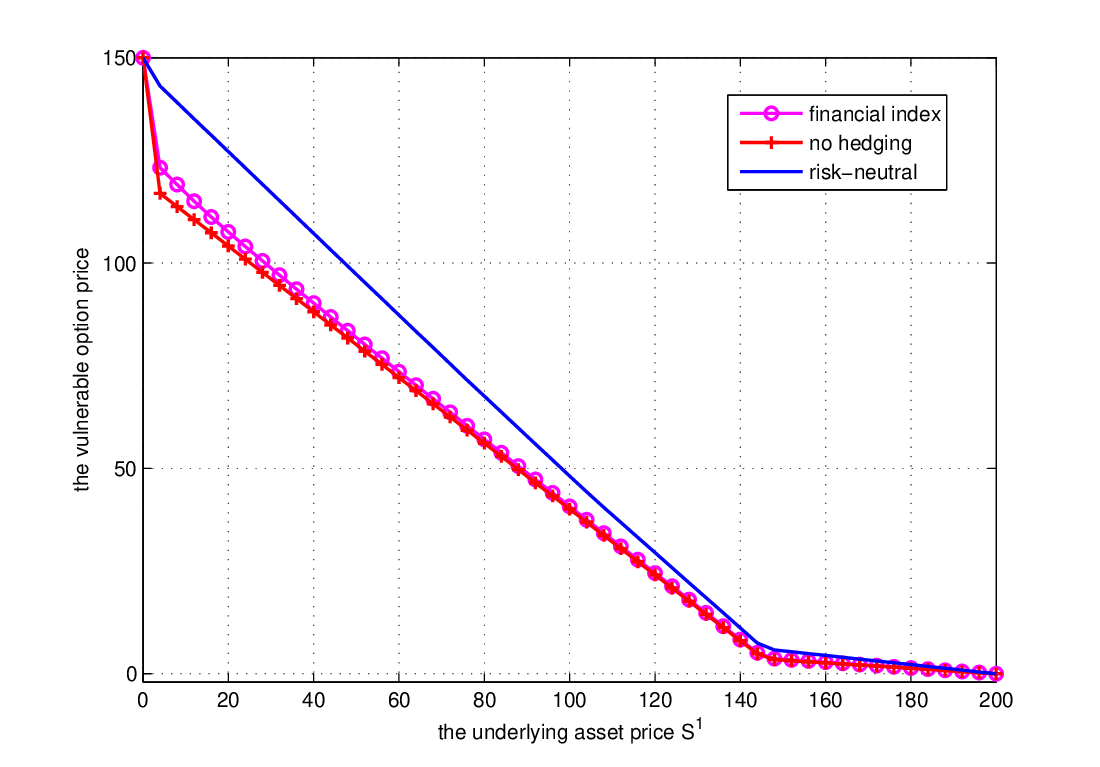}
\end{minipage}
\caption{Vulnerable option price against the underlying asset price
$\exp(S^1)$. The left panel takes $\exp(S^2)=500$; the right panel
takes $\exp(S^2)=1400$.}\label{Fig-S}
\end{figure}

Figure \ref{Fig-S} shows the vulnerable option price(s) against the
underlying asset price $\exp(S^1)$. The two panels of Figure
\ref{Fig-S} are intended to illustrate a ``close or likely to
default'' scenario (the left panel with $\exp(S^2)=500$ relative to
$L=1000$) and a ``far or unlikely to default'' scenario (the right
panel with $\exp(S_2)=1400$ relative to $L=1000$). In both panels
the risk neutral or complete market price is the highest. As the
underlying asset price becomes very large, all option prices tend to
zero, as the put is worthless, regardless of the default risk. At
$\exp(S^1)=0$, in the right panel, the option price is equal to the
option strike $K=150$. In the left panel, the option price is lower
due to the risk of counterparty default. As $S^1$ increases, all
option prices decrease, as the moneyness of the put changes. When
$\exp(S^1)$ is close to zero, we see a dramatic drop in the utility
indifference prices (relative to the risk neutral prices) due to the
risk aversion towards unhedgeable price and default risks. Recall
that since the underlying asset and firm's assets are positively
correlated, default risk becomes more important for low values of
the underlying asset. The price drop is much more significant in the
left panel (and in this extreme case, the option price drops down to
zero if no hedging can be carried out), where the likelihood of
counterparty default is higher. The option holder's risk aversion
causes the utility indifference prices to lie below the risk neutral
price (in each default scenario) with the relative discount to the
risk neutral price being much greater in the left panel where
default is more likely. Assuming the holder can hedge in the
financial index, there is a drop of around 75\% from the
risk-neutral price to the utility indifference price. In the right
panel, where the likelihood of default is relatively low, the
difference between the utility indifference price(s) and the
risk-neutral price is not as dramatic, and is at most around 20\% of
the risk-neutral price. We also see that the ability to hedge in the
correlated financial index (versus no hedging at all) is more
important when the default risk is higher (in the left panel).

\begin{figure}[!htb]
\centering\includegraphics[width=0.55\textwidth]
{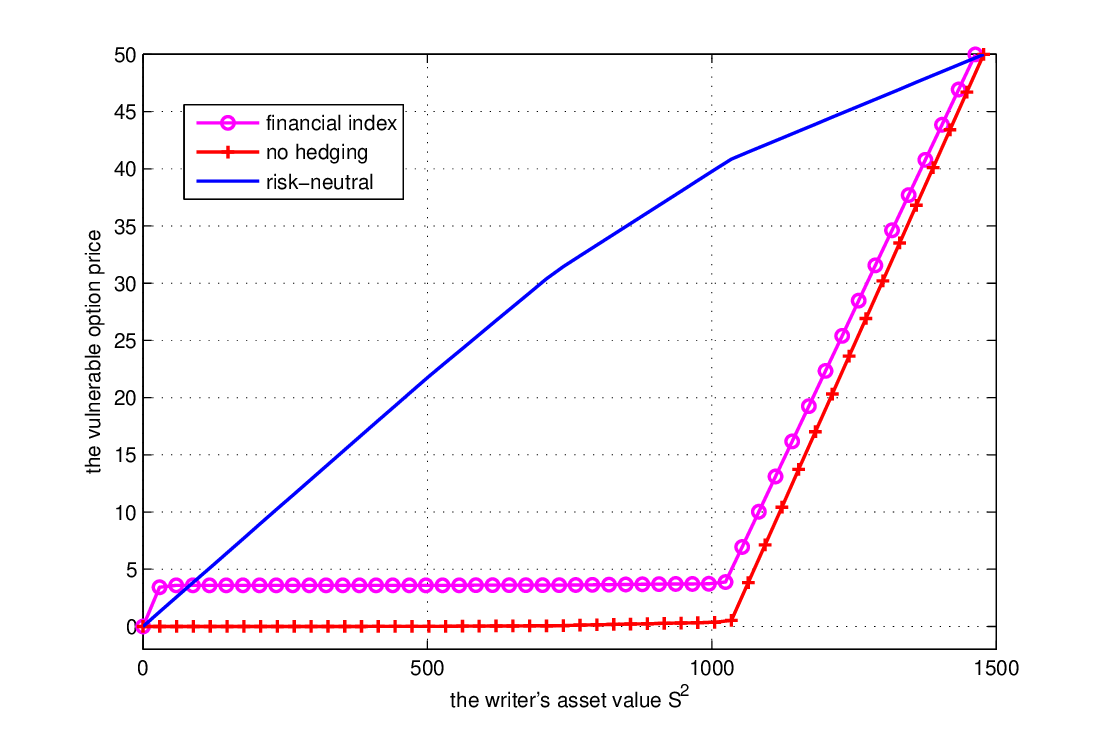} \caption {Vulnerable option price against
the writer's asset value $\exp(S^2)$.} \label{Fig-V}
\end{figure}

Figure \ref{Fig-V} displays the impact of the option writer's asset
value $\exp(S^2)$ on the option price for a fixed asset price
$\exp(S^1)$. We see a dramatic difference in the behavior of the
risk neutral price and the utility indifference prices. Under risk
neutrality, the option price increases smoothly with $\exp(S^2)$.
However, under utility indifference, the prices are low and do not
change much with values of $\exp(S^2)$ below the default trigger of
$L=1000$. This is despite the put being in-the-money. As $\exp(S^2)$
increases beyond the default level, the likelihood of default
diminishes, and the utility indifference prices start to increase
with $\exp(S^2)$. Note that the utility price is not always below
the risk neutral price. Although Proposition \ref{Asy2} tells us
that the risk neutral price is obtained as a limiting case of the
utility indifference price, it requires condition (\ref{CAPM}) to
hold.

\begin{figure}
\begin{minipage}[t]{0.5\linewidth}
\centering
\includegraphics[scale=0.31]{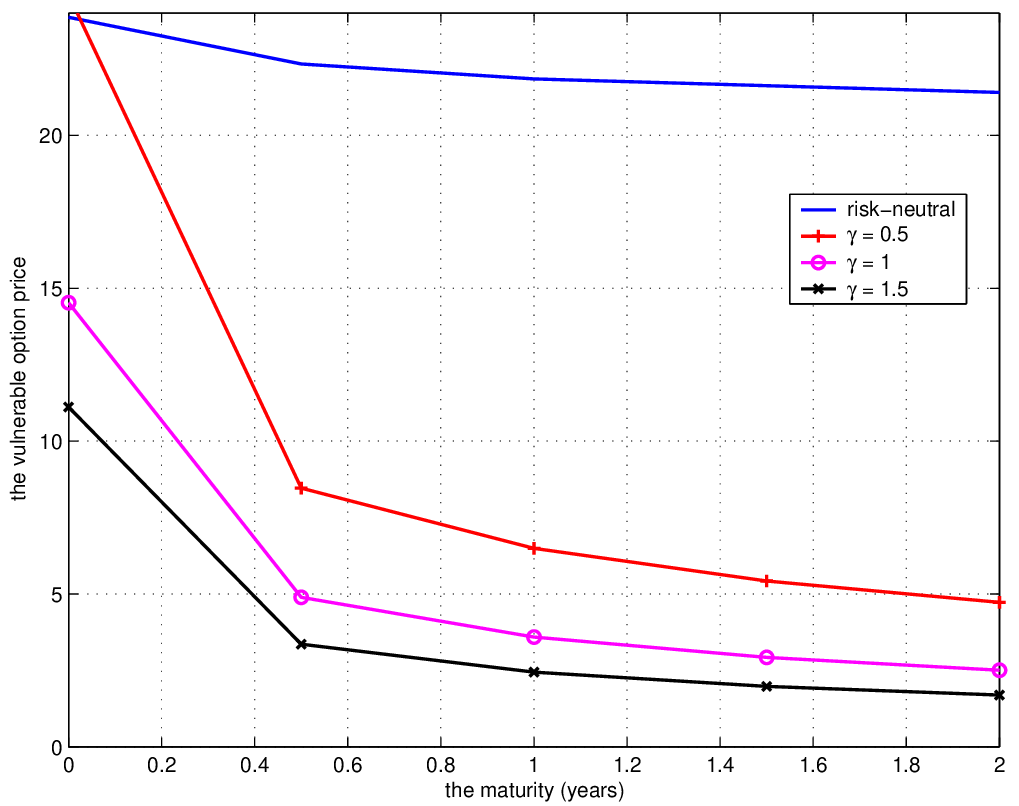}
\end{minipage}
\begin{minipage}[t]{0.5\linewidth}
\centering
\includegraphics[scale=0.35]{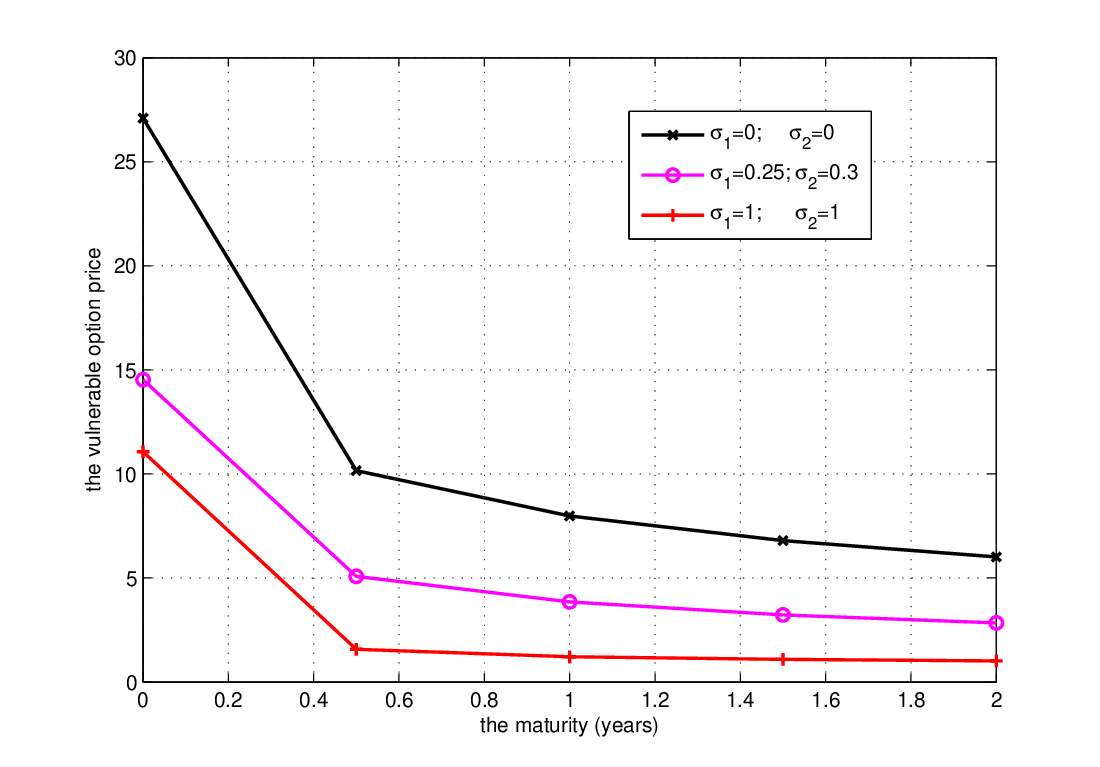}
\end{minipage}
\caption{Impact of risk aversion and correlation. The left panel
gives the option price against maturity for various risk aversion
parameters $\gamma$. The right panel gives the price against various
correlation parameters. We set $\mu_1=0.1$ and $\mu_2=0.06$ to
satisfy the parameter restriction of Proposition \ref{Asy2}.}
\label{Fig-gamma}
\end{figure}

Figure \ref{Fig-gamma} compares how the vulnerable option price
changes with risk aversion parameter $\gamma$, and the idiosyncratic
volatilities $\sigma_1, \sigma_2$. The left panel plots vulnerable
option prices against maturity $T$ for various values of risk
aversion $\gamma$. We see that the more risk averse the option
holder is, the less she will pay for the option, consistent with
Proposition \ref{Asy1}. The other observation is that option prices
for a fixed $\gamma$ are decreasing with maturity $T$. The risk
neutral price is also decreasing with $T$, albeit very gradually.
This is in contrast to risk neutral prices for non-default European
put options which will increase in $T$ provided there are no
dividends. The reason is that there is a tradeoff between price and
default risk. If the maturity is longer, there is more chance for
both $S^1$ and $S^2$ to fall - $S^1$ falling means the put is more
valuable, but $S^2$ falling increases the default risk. For the
parameters considered, the default risk is the dominant factor and
thus the option price decreases with $T$. This is also in contrast
to the call option, where Klein \cite{Klein} reports that the risk
neutral price increases with maturity.

Recall that we do not expect price monotonicity in terms of the
correlations, except in the situation outlined in Proposition
\ref{Asy1}. Here we give an example of prices for various values of
the idiosyncratic volatilities $\sigma_1, \sigma_2$. The left panel
sets parameters to be $\mu_1=0.1$ and $\mu_2=0.06$ to satisfy the
CAPM restriction on Sharpe ratios. If $\sigma_1=\sigma_2=0$, then we
have $\rho_{1 2}=1$, $\rho_{1 P}=0.8$ and $\rho_{2 P}=0.8$.
Similarly if $\sigma_1=0.25$ and $\sigma_2=0.3$ then $\rho_{1
2}=0.4$, $\rho_{1 P}=0.6$ and $\rho_{2 P}=0.4$. Finally, if
$\sigma_1=\sigma_2=1$ then $\rho_{1 2}=0.06$, $\rho_{1 P}=0.2$ and
$\rho_{2 P}=0.16$. We see that as $\sigma_1, \sigma_2$ increase, the
utility indifference price falls. Correspondingly, as the
correlations $\rho_{1 2}, \rho_{1 P}, \rho_{2 P}$ increase, the
option price rises.

%
%
%

\appendix
\section{Proof of Theorem \ref{Markov_lemma_111}}\label{appendix}

\subsection{Derivation of Pricing PDE (\ref{PDE for n nontraded assets})}

The proof is an extension of Theorem 2.3 of Henderson and Liang
\cite{HL}, where we considered the special case $R=1$ in a
non-Markovian setting. We first transform the optimal portfolio
problem (\ref{optm1}) into a risk-sensitive control formulation, and
derive a quadratic BSDE representation for its value process and the
associated optimal trading strategy. The problem (\ref{optm1}) can
be reformulated as
\begin{align*}
&-e^{-\gamma(x-\mathfrak{C}^{\lambda}_t)}\essinf_{\pi\in\mathcal{A}_{\mathcal{F}}[t,T]}
E^{\mathbf{P}}\left[e^{-\gamma\left( \mathbf{1}_{\{t<\tau\leq
T\}}\left(X_{\tau}^{0}(\pi)+R\mathfrak{C}^{\lambda}(\mathcal{S}_{\tau},\tau)\right)+\mathbf{1}_{\{\tau>
T\}}\left(X_T^{0}(\pi)+\lambda g(\mathcal{S}_{T})\right)
 \right)}|\mathcal{G}_t\right]\\
=&-e^{-\gamma(x-\mathfrak{C}^{\lambda}_t)}\exp\left\{-\gamma\esssup_{\pi\in\mathcal{A}_{\mathcal{F}}[t,T]}
\frac{-1}{\gamma}\ln E^{\mathbf{P}}\left[\cdots|\mathcal{G}_t\right]
\right\}.
\end{align*}

For any $\pi\in\mathcal{A}_{\mathcal{F}}[t,T]$, by using the
distribution property of the random time $\tau$ (see Chapter 8 of
Bielecki and Rutkowski \cite{Bielecki-text} for example), we
calculate the above conditional expectation, which is
$\mathbf{1}_{\{\tau>t\}}\bar{Y}_t(\pi)$ with
\begin{align*}
\bar{Y}_t(\pi)=&\
E^{\mathbf{P}}\left[\int_t^Ta(\mathcal{S}_s,s)e^{-\int_t^{s}a(\mathcal{S}_u,u)du
-\gamma X_s^{0}(\pi)-\gamma
R\mathfrak{C}^{\lambda}(\mathcal{S}_s,s)}ds\right.\\
&\left.+\ e^{-\int_t^Ta(S_s,s)ds-\gamma X_T^{0}(\pi)-\gamma\lambda
g(\mathcal{S}_T)}|\mathcal{F}_t \right].
\end{align*}

Therefore, in order to solve the optimal portfolio problem
(\ref{optm1}), we only need to solve the following risk-sensitive
control problem:
\begin{equation}\label{risk-control}
Y_t=\esssup_{\pi\in\mathcal{A}_{\mathcal{F}}[t,T]}Y_t(\pi)=\esssup_{\pi\in\mathcal{A}_{\mathcal{F}}[t,T]}
\frac{-1}{\gamma}\ln\bar{Y}_t(\pi),
\end{equation}
and then the value process of (\ref{optm1}) is given by
\begin{equation}\label{optm1_Sol}
-e^{-\gamma(x-\mathfrak{C}^{\lambda}_t)}\times
\mathbf{1}_{\{\tau>t\}}e^{-\gamma Y_t}.
\end{equation}

In the following, we will first work out the BSDE representation for
$Y_t(\pi)$, and then obtain the BSDE representation for $Y_t$ by the
comparison principle. For any
$\pi\in\mathcal{A}_{\mathcal{F}}[0,T]$, we change the probability
measure from $\mathbf{P}$ to $\mathbf{Q}(\pi)$ by defining
$$\frac{d\mathbf{Q}(\pi)}{d\mathbf{P}}=\mathcal{E}(N(\pi))
=\mathcal{E}\left(-\int_0^{\cdot}\gamma\pi_s(\bar{\sigma}_PdW_s^{n+1}+\sigma_PdW_s^{n+2})\right),$$
where $\mathcal{E}(\cdot)$ is the Dol\'eans-Dade exponential. By
Girsanov's theorem,
$\mathcal{B}(\pi)=\mathcal{W}-\langle\mathcal{W},N(\pi)\rangle$ is
the Brownian motion under the new probability measure
$\mathbf{Q}(\pi)$. With the new Brownian motion
$\mathcal{B}(\pi)=(B^1(\pi),\dots,B^{n+2}(\pi))$, the logarithm
price processes of the non-traded assets
$\mathcal{S}=(S^1,\cdots,S^n)$ satisfy
\begin{equation}\label{dynamics of nontraded assets_22}
{dS_t^{i}}=\left(\mu_i(\mathcal{S}_t,t)-\gamma\bar{\sigma}_i(\mathcal{S}_t,t)\bar{\sigma}_P\pi_t\right)dt
+\sigma_{i}(\mathcal{S}_t,t)dB_t^i(\pi)+\bar{\sigma}_{i}(\mathcal{S}_t,t)dB_t^{n+1}(\pi),
\end{equation}
and $\bar{Y}_t(\pi)$ becomes
\begin{align*}
\bar{Y}_t(\pi)=&\
E^{\mathbf{P}}\left[\int_t^T\mathcal{E}(N(\pi))_se^{-\int_t^s\rho_u(\pi)du}
a(\mathcal{S}_s,s)e^{-\gamma
R\mathfrak{C}^{\lambda}(\mathcal{S}_s,s)}ds\right.\\
&\ \ \ \ \ \ \ \
+\left.\mathcal{E}(N(\pi))_Te^{-\int_t^T\rho_u(\pi)du}
e^{-\gamma\lambda
g(\mathcal{S}_T)}|\mathcal{F}_t \right]\\
=&\ E^{\mathbf{Q}(\pi)}\left[\int_t^Te^{-\int_t^s\rho_u(\pi)du}
a(\mathcal{S}_s,s)e^{-\gamma
R\mathfrak{C}^{\lambda}(\mathcal{S}_s,s)}ds+
e^{-\int_t^T\rho_u(\pi)du} e^{-\gamma\lambda
g(\mathcal{S}_T)}|\mathcal{F}_t \right],
\end{align*}
with the stochastic discount factor $\rho_u(\pi)$:
$$\rho_u(\pi)=a(\mathcal{S}_u,u)+\gamma\mu_P\pi_u-\frac{\gamma^2}{2}\left(\sigma_P^2+\bar{\sigma}_P^2\right)|\pi_u|^2.$$

By the martingale representation theorem, there exists an
$\mathbb{R}^{n+1}$-valued predictable process
$\bar{Z}(\pi)=(\bar{Z}^1(\pi),\dots,\bar{Z}^{n+1}(\pi))$ such that
\begin{equation}
\bar{Y}_t(\pi)=e^{-\gamma\lambda
g(\mathcal{S}_T)}+\int_t^T\left[a(\mathcal{S}_s,s)e^{-\gamma
R\mathfrak{C}^{\lambda}(\mathcal{S}_s,s)}-\rho_s(\pi)\bar{Y}_s(\pi)\right]ds-\sum_{i=1}^{n+1}\int_t^T\bar{Z}^i_s(\pi)dB_s^i(\pi).
\end{equation}
From (\ref{risk-control}),
$Y_t(\pi)=-\frac{1}{\gamma}\ln\bar{Y}_t(\pi)$. It\^o's formula then
implies that
\begin{equation}
Y_t(\pi)=\lambda
g(\mathcal{S}_T)+\int_t^T\left[\frac{\rho_s(\pi)}{\gamma}-\frac{a(\mathcal{S}_s,s)}{\gamma}e^{\gamma(1-R)\mathfrak{C}^{\lambda}(\mathcal{S}_s,s)}-\frac{\gamma}{2}|Z_s(\pi)|^2\right]ds-\sum_{i=1}^{n+1}\int_t^TZ_s^{i}(\pi)dB_s^i(\pi),
\end{equation}
where
$Z_t^i(\pi)=\frac{-1}{\gamma}{\bar{Z}_t^{i}(\pi)}/{\bar{Y}_t(\pi)}$.
Equivalently, under the original probability measure $\mathbf{P}$,
we write
\begin{align}\label{auxiliaryBSDE}
Y_t(\pi)=&\ \lambda
g(\mathcal{S}_T)+\int_t^T\left[\frac{\rho_s(\pi)}{\gamma}-\gamma\bar{\sigma}_PZ_s^{n+1}(\pi)\pi_s-\frac{a(\mathcal{S}_s,s)}{\gamma}e^{\gamma(1-R)\mathfrak{C}^{\lambda}(\mathcal{S}_s,s)}-\frac{\gamma}{2}|Z_s(\pi)|^2\right]ds\notag\\
&-\sum_{i=1}^{n+1}\int_t^TZ_s^{i}(\pi)dW_s^i.
\end{align}

We notice that (\ref{auxiliaryBSDE}) is a quadratic BSDE with
bounded terminal data and coefficients, whose existence and
uniqueness is guaranteed by Theorems 2.3 and 2.6 of Kobylanski
\cite{Kobylanski}. Moreover, the comparison principle holds for
(\ref{auxiliaryBSDE}). Let
$\pi^1,\pi^2\in\mathcal{A}_{\mathcal{F}}[0,T]$ such that
$$\frac{\rho_s(\pi^1)}{\gamma}-\gamma\bar{\sigma}_Pz\pi^1_s\geq
\frac{\rho_s(\pi^2)}{\gamma}-\gamma\bar{\sigma}_Pz\pi^2_s$$ for
$z\in\mathbb{R}^{n+1}$. Then $Y_t({\pi^1})\geq Y_t({\pi^2})$.

We claim that the value process of our risk-sensitive control
problem (\ref{risk-control}) is given by the solution of the
following quadratic BSDE:
\begin{align}\label{quadraticBSDE}
Y_t=&\ \lambda
g(\mathcal{S}_T)+\int_t^T\left[\frac{a(\mathcal{S}_s,s)}{\gamma}(1-e^{\gamma(1-R)\mathfrak{C}^{\lambda}(\mathcal{S}_s,s)})+\frac{\left(\mu_P-\gamma\bar{\sigma}_PZ^{n+1}_s\right)^2}{2\gamma[(\sigma_P)^2+(\bar{\sigma}_P)^2]}-\frac{\gamma}{2}|Z_s|^2\right]ds\notag\\
&-\sum_{i=1}^{n+1}\int_t^TZ_s^{i}dW_s^i,
\end{align}
with the associated optimal portfolio:
$$\pi_t^{*,\lambda}=\frac{\bar{\vartheta}^P}{\gamma\bar{\sigma}_P}-\frac{\bar{\kappa}_P}{\bar{\sigma}_P}Z_t^{n+1}.$$
Indeed, by Theorems 2.3 and 2.6 of Kobylanski \cite{Kobylanski}, the
quadratic BSDE (\ref{quadraticBSDE}) admits a unique bounded
solution $Y$ with its martingale representation part
$Z=(Z^1,\dots,Z^{n+1})$. To prove the claim, we notice that for any
$\pi\in\mathcal{A}_{\mathcal{F}}[t,T]$,
$$\frac{\rho_s(\pi)}{\gamma}-\gamma\bar{\sigma}_PZ_s^{n+1}\pi_s\leq
\frac{a(\mathcal{S}_s,s)}{\gamma}+\frac{\left(\mu_P-\gamma\bar{\sigma}_PZ^{n+1}_s\right)^2}{2\gamma[(\sigma_P)^2+(\bar{\sigma}_P)^2]},$$
and for $\pi=\pi^{*,\lambda}$, the equality holds:
$$\frac{\rho_s(\pi^{*,\lambda})}{\gamma}-\gamma\bar{\sigma}_PZ_s^{n+1}\pi_s^{*,\lambda}=
\frac{a(\mathcal{S}_s,s)}{\gamma}+\frac{\left(\mu_P-\gamma\bar{\sigma}_PZ^{n+1}_s\right)^2}{2\gamma[(\sigma_P)^2+(\bar{\sigma}_P)^2]}.$$
Then the claim follows from the comparison principle for
(\ref{auxiliaryBSDE}).

The optimization problem (\ref{optm2}) is a special case of
(\ref{optm1}) with $\lambda=0$, whose value process is given by
$-e^{-\gamma x}\times
\mathbf{1}_{\{\tau>t\}}e^{-\frac{\bar{\theta}^P}{2}(T-t)},$ and the
optimal control on the event $\{\tau>t\}$ is given by
$\pi^{*,0}_t=\frac{\bar{\vartheta}^P}{\gamma\bar{\sigma}_P}$.

By Definition \ref{definition1}, the utility indifference price
$\mathfrak{C}_t^{\lambda}$ is such that
$$-e^{-\gamma(x-\mathfrak{C}_t^{\lambda})}\times
\mathbf{1}_{\{\tau>t\}}e^{-\gamma Y_t}=-e^{-\gamma x}\times
\mathbf{1}_{\{\tau>t\}}e^{-\frac{\bar{\theta}^P}{2}(T-t)},$$ from
which we obtain
$$\mathfrak{C}_t^{\lambda}=\mathbf{1}_{\{\tau>t\}}\left(Y_t-\frac{\bar{\theta}^P}{2\gamma}(T-t)\right).$$
The hedging strategy ${\pi}^{*,\lambda}_t-\pi_t^{*,0}$ on the event
$\{\tau>t\}$ is given as
$-\frac{\bar{\kappa}_P}{\bar{\sigma}_P}Z_t^{n+1}$.

Finally, the pricing PDE (\ref{PDE for n nontraded assets}) is
obtained by an application of nonlinear Feynman-Kac representation
(see Section 3 of \cite{Kobylanski}) by noticing that
$Y_t-\frac{\bar{\theta}^P}{2\gamma}(T-t)=\mathfrak{C}^{\lambda}(\mathcal{S}_t,t)$
and
$$Z_t^i=\sigma_i(\mathcal{S}_t,t)\partial_{s_i}\mathfrak{C}^{\lambda}(\mathcal{S}_t,t)\ \ \text{for}\ \ i=1,\dots,n;\ \ \
Z_t^{n+1}=\sum_{i=1}^n\bar{\sigma}_i(\mathcal{S}_t,t)\partial_{s_{i}}\mathfrak{C}^{\lambda}(\mathcal{S}_t,t).$$

\subsection{Well posedness of the Pricing PDE (\ref{PDE for n nontraded assets})}
\label{sec-boundedsolution}

The existence and uniqueness of bounded solutions to (\ref{PDE for n
nontraded assets}) are easily obtained from \cite{Kobylanski}. We
next derive explicit bounds for the solution
$\mathfrak{C}^{\lambda}(\mathbf{s},t)$, which are used in
Proposition \ref{Asy1}.

It is obvious that the solution is nonnegative:
$\mathfrak{C}^{\lambda}(\mathbf{s},t)\geq 0$. Let
$\mathfrak{C}^{\lambda}(\mathbf{s},t;I)$ be the solution of the
pricing PDE with the full recovery $R=1$, so it satisfies the
following PDE with quadratic gradients:
\begin{eqnarray}\label{PDE for n nontraded assets1}
    \left\{
    \begin{array}{ll}
    -\partial_t\mathfrak{C}^{\lambda}(\mathbf{s},t;I)-(\mathbf{L}^1+
    \mathbf{L}^2)\mathfrak{C}^{\lambda}(\mathbf{s},t;I)=0,
    \\
    \mathfrak{C}^{\lambda}(\mathbf{s},T;I)=\lambda g(\mathbf{s}).
    \end{array}\right.
\end{eqnarray}
Note that $-(\partial_t+\mathbf{L}^1+
    \mathbf{L}^2+\mathbf{L}^3)\mathfrak{C}^{\lambda}(\mathbf{s},t;I)\geq 0$, since
$$\mathbf{L}^3\mathfrak{C}^{\lambda}(\mathbf{s},t;I)\leq \frac{a(\mathbf{s})}{\gamma}\left[1-e^{\gamma(1-R)\mathfrak{C}^{\lambda}(\mathbf{s},t;I)
}\right]=0,$$ so that $\mathfrak{C}^{\lambda}(\mathbf{s},t;I)$ is a
supersolution to (\ref{PDE for n nontraded assets}). On the other
hand, $\mathfrak{C}^{\lambda}(\mathbf{s},t)$ is the (sub)solution to
(\ref{PDE for n nontraded assets}), and
$\mathfrak{C}^{\lambda}(\mathbf{s},T;I)=\mathfrak{C}^{\lambda}(\mathbf{s},T)$.
By the comparison principle, we conclude that
$\mathfrak{C}^{\lambda}(\mathbf{s},t)\leq
\mathfrak{C}^{\lambda}(\mathbf{s}, t;I)$ on
$\mathbb{R}^{n}\times[0,T]$.

The solution $\mathfrak{C}^{\lambda}(\mathbf{s},t;I)$ to PDE
(\ref{PDE for n nontraded assets1}) is interpreted as the utility
indifference price without intertemporal default, and we claim that
$\mathfrak{C}^{\lambda}(\mathbf{s},t;I)$ is bounded from above by
$\bar{\mathfrak{C}}^{\lambda}(\mathbf{s},t;I)$, which is the
solution to PDE (\ref{PDE for n nontraded assets1}) with $\gamma=0$,
that is, $\bar{\mathfrak{C}}^{\lambda}(\mathbf{s},t;I)$ satisfies
the following linear PDE:
\begin{eqnarray}\label{PDE for n nontraded assets2}
    \left\{
    \begin{array}{ll}
    -\displaystyle\partial_t\bar{\mathfrak{C}}^{\lambda}(\mathbf{s},t;I)-\mathbf{L}^1\bar{\mathfrak{C}}^{\lambda}(\mathbf{s},t;I)
    +\sum_{i=1}^n\bar{\vartheta^P}\bar{\sigma}_i\partial_{s_i}\bar{\mathfrak{C}}^{\lambda}(\mathbf{s},t;I)
    =0,
    \\
    \bar{\mathfrak{C}}^{\lambda}(\mathbf{s},T;I)=\lambda
    g(\mathbf{s}).
    \end{array}\right.
\end{eqnarray}
Indeed, note that $-(\partial_t+\mathbf{L}^1+
    \mathbf{L}^2)\bar{\mathfrak{C}}^{\lambda}(\mathbf{s},t;I)\geq 0$, since the terms involving
$\gamma$ in $\mathbf{L}^2$ can be regrouped as (omitting the
arguments in $\sigma_i$ and $\bar{\sigma}_i$)
\[
-\frac{\gamma}{2}\sum_{i=1}^n\sigma_i^2\left(\partial_{s_i}\bar{\mathfrak{C}}^{\lambda}(\mathbf{s},t;I)\right)^2
 -\frac{\gamma}{2}(1-\bar{\kappa}^P) \left(\sum_{i=1}^{n} \bar{\sigma}_i \partial_{s_i}\bar{\mathfrak{C}}^{\lambda}(\mathbf{s},t;I)\right)^2\leq 0
\]
so that $\bar{\mathfrak{C}}^{\lambda}(\mathbf{s},t;I)$ is a
supersolution to (\ref{PDE for n nontraded assets1}). On the other
hand, $\mathfrak{C}^{\lambda}(\mathbf{s},t;I)$ is the (sub)solution
to (\ref{PDE for n nontraded assets1}), and
$\bar{\mathfrak{C}}^{\lambda}(\mathbf{s},T;I)=
\mathfrak{C}^{\lambda}(\mathbf{s},T;I)$. The comparison principle
implies that  $\mathfrak{C}^{\lambda}(\mathbf{s},t;I)\leq
\bar{\mathfrak{C}}^{\lambda}(\mathbf{s},t;I)$ on
$\mathbb{R}^n\times[0,T]$.

It is well known that the linear PDE (\ref{PDE for n nontraded
assets2}) admits a unique bounded solution since the terminal data
$g(\cdot)$ is bounded. Hence, there exists a constant $K$ such that
$$0\leq \mathfrak{C}^{\lambda}(\mathbf{s},t)\leq \mathfrak{C}^{\lambda}(\mathbf{s},t;I)\leq
\bar{\mathfrak{C}}^{\lambda}(\mathbf{s},t;I) \leq K.$$

Finally, we show that $\mathfrak{C}^{\lambda}(\mathbf{s},t)$ is
Lipschitz continuous in $\mathbf{s}$. This in turn gives us the
uniform boundedness of ${\pi}^{*,\lambda}$ and therefore the hedging
strategy $\pi^{*,\lambda}-\pi^{*,0}$. Indeed, since $a(\cdot,\cdot)$
is bounded and $\mathfrak{C}^{\lambda}(\mathbf{s},t)\in[0,K]$,
$\mathbf{L}^3(x)$ is Lipschitz continuous in $x\in[0,K]$. On the
other hand, $\mathbf{L}^2$ has at most quadratic growth in
gradients. We can apply Theorem 2.9 of Delarue \cite{MR2053051} to
conclude that the solution $\mathfrak{C}^{\lambda}(\mathbf{s},t)$
indeed has bounded gradients:
$\partial_{\mathbf{s}}\mathfrak{C}^{\lambda}(\mathbf{s},t)$ is
uniformly bounded.

%

\newpage

\end{document}